\documentclass[useAMS,usenatbib]{mn2e}
\usepackage[]{txfonts}
\usepackage{subfigure}
\usepackage{epsfig}
\graphicspath{{/home/av/metzger/timokhin/graphs/}{/home/av/metzger/timokhin/graphs/per02.0/}}
\usepackage{url}
\def\simless{\mathbin{\lower 3pt\hbox
{$\rlap{\raise 5pt\hbox{$\char'074$}}\mathchar"7218$}}}   
\def\simmore{\mathbin{\lower 3pt\hbox
{$\rlap{\raise 5pt\hbox{$\char'076$}}\mathchar"7218$}}}   
\newcommand       \apj          {ApJ}
\newcommand       \apjl         {ApJL}
\newcommand       \aap          {A\&A}
\newcommand       \nat          {Nature}
\newcommand       \mnras        {MNRAS}

\newcommand       \prd      {Phys.~Rev.~D.~}

\newcommand       \pasj   {PASJ}
\newcommand      \apjs {ApJ Supplements}
\newcommand      \na {New Astronomy}
\newcommand      \physrep {Physics Reports}

\def\simlt{\mathrel{\hbox{\rlap{\hbox{\lower4pt\hbox{$\sim$}}}\hbox{$<$}}}}
\def\simgt{\mathrel{\hbox{\rlap{\hbox{\lower4pt\hbox{$\sim$}}}\hbox{$>$}}}}
\def\lesssim{\mathrel{\hbox{\rlap{\hbox{\lower4pt\hbox{$\sim$}}}\hbox{$<$}}}}

\def\gtrsim{\mathrel{\hbox{\rlap{\hbox{\lower4pt\hbox{$\sim$}}}\hbox{$>$}}}}

\newcommand{\p}{\partial}
\newcommand{\be}{\begin{eqnarray}}
\newcommand{\ee}{\end{eqnarray}}

\newcommand{\De}{\Delta}
\newcommand{\eps}{\varepsilon}
\renewcommand{\ln}{{\rm ln}}

\newcommand{\pt}[3]{\left(\frac{\p {#1}}{\p {#2}}\right)_{#3}}

\title{Neutrino-heated winds from rotating proto-magnetars}

\author[]{Andrey~D.~Vlasov$^1\thanks{E-mail: adv2110@columbia.edu}$, Brian~D.~Metzger$^{1}\thanks{E-mail: bmetzger@phys.columbia.edu}$, Todd A.~Thompson$^{2}\thanks{E-mail: thompson@astronomy.ohio-state.edu}$\\
$^{1}$Columbia Astrophysics Laboratory, Columbia University, New York, NY, 10027, USA\\
$^{2}$Department of Astronomy and Center for Cosmology and AstroParticle Physics, The Ohio State University, Columbus, Ohio 43210, USA}

\begin{document}
\date{Received / Accepted}
\pagerange{\pageref{firstpage}--\pageref{lastpage}} \pubyear{2014}

\maketitle

\label{firstpage}
\begin{abstract}
We calculate the steady-state properties of neutrino-driven winds from strongly magnetized, rotating proto-neutron stars (`proto-magnetars') under the assumption that the outflow geometry is set by the force-free magnetic field of an aligned dipole.  Our goal is to assess proto-magnetars as sites of $r$-process nucleosynthesis and gamma-ray burst engines.  One dimensional solutions calculated along flux tubes corresponding to different polar field lines are stitched together to determine the global properties of the flow at a given neutrino luminosity and rotation period. 
Proto-magnetars with rotation periods of $P \sim 2-5$ ms are shown to produce outflows more favorable for the production of third-peak $r$-process nuclei due to their much shorter expansion times through the seed nucleus formation region, yet only moderately lower entropies, as compared to normal spherical PNS winds.  Proto-magnetars with moderately rapid birth periods $P \sim 3-5$ ms may thus represent a promising Galactic $r$-process site which is compatible with a variety of other observations, including the recent discovery of possible magnetar-powered supernovae in metal poor galaxies.  We also confirm previous results that the outflows from proto-magnetars with $P \sim 1-2$ ms can achieve maximum Lorentz factors $\Gamma_{\rm max} \sim 100-1000$ in the range necessary to power gamma-ray bursts (GRBs).  The implications of GRB jets with a heavy nuclei-dominated composition as sources of ultra-high energy cosmic rays are also addressed.  
\end{abstract} 
  
\section{Introduction} 
\label{sec:intro}
All neutron stars are born as hot `proto-neutron stars' (PNSs), which radiate their gravitational binding energy and lepton number in neutrinos over the first $\sim 10-100$ seconds of their life \citep{Burrows&Lattimer86}.  A small fraction of this luminosity is absorbed by matter in the atmosphere just above the PNS surface, powering an outflow of mass known as the `neutrino-driven wind' (e.g.~\citealt{Duncan+86}; \citealt{Qian&Woosley96}).  Initial interest in neutrino winds arose due to their potential as a primary source of rapid neutron capture ($r$-process) nucleosynthesis (e.g.~\citealt{Meyer+92}; \citealt{Takahashi+94}; \citealt{Woosley+94}).  Neutrino-heated outflows are, however, also of great importance in other contexts, such as their role in determining the maximum Lorentz factor (baryon loading) and nuclear composition of gamma-ray burst outflows (e.g.~\citealt{Thompson+04}; \citealt{Bucciantini+07}; \citealt{Metzger+07,Metzger+08a,Metzger+11a}; \citealt{Barzilay&Levinson08}; \citealt{Lei+13}).

Past studies of neutrino-driven winds have been primarily focused on spherically symmetric, non-rotating winds accelerated purely by thermal pressure (\citealt{Kajino+00}; \citealt{Sumiyoshi+00}; \citealt{Otsuki+00}; \citealt{Thompson+01}; \citealt{Arcones&Montes11}; \citealt{Roberts+10}; \citealt{Roberts+12}; \citealt{MartinezPinedo+12}; \citealt{Fischer+12}).  This body of work has led to the conclusion that normal PNS winds fail to achieve the conditions necessary for nucleosynthesis to reach the third $r$-process peak at mass number $A \sim 195$.  The latter requires an outflow with a combination of high entropy $S$, short expansion timescale $t_{\rm exp}$, and low electron fraction $Y_e$ (\citealt{Hoffman+97}) as it passess through the radii where seed nuclei form.  Proposed mechanisms to move beyond the standard scenario in order to achieve $r$-process success include postulating additional sources of heating (e.g. damping of convectively-excited waves; \citealt{Suzuki+Nagataki05}; \citealt{Metzger+07}) or by resorting to extreme parameters, such as massive $\gtrsim 2.2M_{\odot}$ neutron stars (\citealt{Wanajo13}).

The class of neutron stars known as `magnetars' possess very strong magnetic fields $\gtrsim 10^{14}-10^{15}$ G and account for at least $\sim 10$ per cent of neutron star births (\citealt{Woods&Thompson06}).  Basic estimates show that magnetar-strength fields are dynamically important in PNS winds (\citealt{Thompson03}), even in the first few seconds after core bounce when the neutrino heating and wind kinetic energy are highest.  Strong magnetic fields confine the outflow to only a fraction of the PNS surface threaded by open field lines (\citealt{Thompson03}).  Rotation, when coupled with a strong magnetic field, also provides an additional source of acceleration in the wind via magneto-centrifugal forces \citep{Thompson+04}.  Strong magnetic fields and rapid rotation do not represent separate assumptions if rapid rotation at birth is necessary to produce the strong field via dynamo action (e.g.~\citealt{Thompson&Duncan93}; \citealt{Akiyama+03}). 
 
\citet{Metzger+07} calculated the steady-state structure of rotating magnetized PNS winds in ideal MHD under the assumption of an equatorial (one-dimensional) outflow with an assumed monopolar radial field $B_r \propto 1/r^{2}$.  \citet{Metzger+07} found that strong magnetic fields $\gtrsim 10^{14}-10^{15}$ G and rapid rotation (spin period $P \lesssim $ few ms) act to increase both the power and the mass loss rate of the wind as compared to the spherical case (see also \citealt{Thompson+04}).  The conditions for a successful $r$-process, as quantified by the critical ratio $S^{3}/t_{\rm exp}$ we found to be moderately improved from the spherical case.  However, in some cases of extreme rotation the conditions were in fact {\it worse} because although magneto-centrifugal acceleration increases the outflow velocity (reducing $t_{\rm exp}$), the residence time of matter in the heating region$-$and hence its final entropy$-$was likewise reduced in a compensatory way.  

A successful $r$-process can take place even without high entropy if the asymptotic electron fraction of the outflow $Y_e$ is substantially lower than the equilibrium value $Y_e^{\rm eq} \gtrsim 0.4$ set by neutrino irradiation (eq.~[\ref{eq:Yeeq}]).  This is conceivable if the outflow expands from the surface, where $Y_e$ starts much lower, sufficiently rapidly as the result of centrifugal acceleration to avoid entering such an equilibrium.  In practice, however, both simple analytic estimates ($\S\ref{sec:Ye}$) and numerical simulations show that $Y_e \ll Y_e^{\rm eq}$ is achieved only for extremely rapid rotation ($P \lesssim 1$ ms) near the centrifugal break-up limit (\citealt{Metzger+08a},\citealt{Winteler+12}).  Extreme rotation necessarily increases the total mass ejected per event by a factor $\gtrsim 10-100$ (\citealt{Metzger+08a}), placing such sources in tension with studies of Galactic chemical evolution that suggest more frequent, lower-mass pollution events are more consistent with the abundance distributions of metal poor stars (\citealt{Argast+04}; \citealt{Matteucci+14}; \citealt{Komiya+14}; however, see \citealt{Tsujimoto&Shigeyama14}).  Most Galactic magnetars cannot be born with extremely short spin periods $P \sim$ 1 ms, as their high birth rate would overproduce the Galactic $r$-process \citep{Ripley+14} and the energy of the explosion accompanying their birth would exceed those measured for supernova remnants hosting magnetars \citep{Vink&Kuiper06}.

Here we present calculations of neutrino-heated outflows from magnetized, rotating PNSs in the more realistic geometry of an aligned (axisymmetric) dipole.  Our solutions are calculated under the force-free assumption, namely that mass flows along the magnetic field lines, with the latter prescribed by a self-consistent solution of the cross-field force balance (\citealt{Contopoulos+99}; \citealt{Spitkovsky06}; \citealt{Timokhin06}). A series of one-dimensional solutions calculated along flux tubes corresponding to different polar field lines are then pieced together to determine the global properties of the flow at a given neutrino luminosity and rotation period.    

A primary conclusion of this work is that proto-magnetar outflows are more favorable for the $r$-process than slowly rotating, unmagnetized PNSs by a factor up to $\sim 4$ in $S^{3}/t_{\rm exp}$.  This improvement results largely due to the faster expansion rate of the outflow from centrifugal slinging, which is accompanied by a more modest decrease in entropy than in the equatorial monopole case.  These improved $r$-process conditions are found to extend to spin periods $P \gtrsim 5$ ms, indicating that core collapse events with extremely high angular momentum are not necessary for a marked improvement.  

Beyond the $r$-process, our results have implications for the viability of millisecond proto-magnetars as potential central engines of GRBs (\citealt{Usov92}; \citealt{Wheeler+00}; \citealt{Thompson+04}; \citealt{Metzger+11a}).  A long-standing theoretical mystery is why GRB jets accelerate to bulk Lorentz factors $\Gamma$ in the relatively narrow range of $\sim 100-1000$ (\citealt{Lithwick&Sari01}), as this requires a finely tuned baryon loading (\citealt{Piran99}).  Here we confirm previous results that proto-magnetar outflows with spin periods $P \sim 1-2$ ms in the range necessary to explain the energetics of GRBs come naturally `loaded' with the correct baryon flux to achieve the observed values of $\Gamma$.  Our results also confirm the prediction that the nuclear composition of GRB jets may be dominated heavy nuclei instead of protons (\citealt*{Metzger+11b}). There are some indications that ultra-high energy cosmic rays (UHECRs; \citealt{Waxman95}) may be dominated by heavy nuclei (\citealt{Abraham+10}; however, see e.g. \citealt{Sokolsky&Thomson07}). This would be naturally explained if millisecond magnetar birth is the origin of UHECRs. 

This paper is organized as follows.  In $\S\ref{sec:methods}$ we describe the equations and method of our calculations.  In $\S\ref{sec:results}$ we present the results of our wind solutions.  In $\S\ref{sec:discussion}$ we discuss the implications of our results for $r$-process nucleosynthesis ($\S\ref{sec:rprocess}$) and gamma-ray burst outflows ($\S\ref{sec:GRB}$).  In $\S\ref{sec:conclusions}$ we provide a bulleted summary of our conclusions.  

\section{Methods}
\label{sec:methods}
\begin{figure}
\begin{center}
\includegraphics[width=1.0\linewidth]{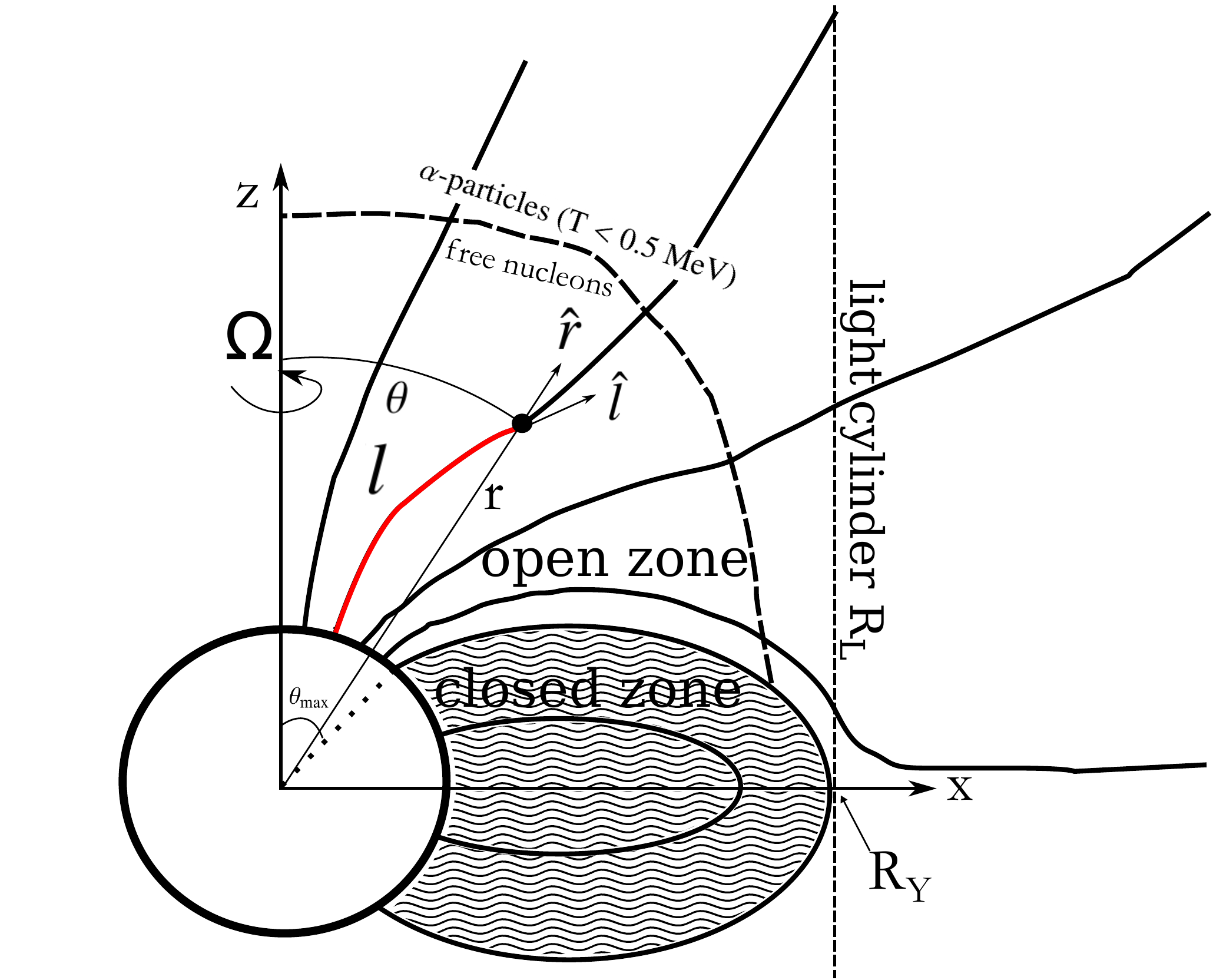}
\caption{Geometry of neutrino-heated winds from magnetized, rotating proto-neutron stars.  Outflows occur from the `open zone',  along field lines with polar angles $\theta < \theta_{\rm max}$, where $\theta_{\rm max} = \sin^{-1}\left[R_{\rm ns}/R_{\rm Y}\right]$ is the angle of the last closed field line, $R_{\rm ns}$ is the PNS radius, $R_{\rm Y} \lesssim R_{\rm L}$ is the radius at which the last closed field line crosses the equator (the `Y point'), $R_{\rm L} = c/\Omega$ is the light cylinder radius, and $\Omega$ is the angular rotation rate of the PNS.  In the illustration we have assumed $R_{\rm Y} = R_{\rm L}$.  The variable $l$ defines the distance measured along the field line from the surface, with the local field line direction denoted as $\hat{l}$ (the toroidal component of which is not shown).  Note that $l$ differs from the spherical radius $r$ measured to the PNS center.  The wind is composed of free nucleons at small radii, which recombine into $\alpha$-particles and heavier seed nuclei once the temperature decreases to $\lesssim 0.5$ MeV several neutron star radii above the surface.}
\label{fig:schematic}
\end{center}
\end{figure}

\subsection{Model Description}

We calculate steady-state outflows along open field lines corresponding to the solutions to the Grad Shafranov equations for an aligned dipole field (\citealt{Timokhin06}).  The steady-state approximation is valid because the properties of the PNS (rotation period $P$, neutrino luminosity $L_{\nu}$, radius $R_{\rm ns}$) change slowly compared to the time for matter to flow to the radii of interest.  In force-free outflows the matter flows along the prescribed field geometry.  This approximation is justified if the pressure $P$ and kinetic energy density $\rho v^2/2$ are much less than magnetic energy density $B^2/8\pi$ (where $\rho$ and $v$ are the density and velocity of the outflow, respectively), a condition we check {\it a posteriori} from our derived solutions.

The simplification that matter moves along the field lines reduces the 2D magneto-hydrodynamics problem to a series of 1D calculations along flux tubes centered around each field line.  Calculations are performed for different initial polar angles $\theta$, with results then averaged or integrated (when appropriate) over the open magnetosphere.  The latter includes angles $\theta < \theta_{\rm max}$, where $\theta_{\rm max} = \sin^{-1}\sqrt{R_{\rm ns}/R_{\rm Y}}$ is the angle of the last closed field line, $R_{\rm Y}$ is the radius at which the last closed field line crosses the equator (the `Y point'), and $R_{\rm L} = c/\Omega = cP/2\pi$ is the light cylinder radius.  Even if the force-free approximation is valid in the wind itself, the location of the Y point, and hence the size of the open zone, can be affected by the pressure of the closed zone (\citealt{Mestel&Spruit87}; $\S\ref{sec:magnmodel}$, $\S\ref{sec:forcefree}$).  The standard spherically-symmetric case is accomodated within our framework by assuming a monopole magnetic field $B_r \propto 1/r^{2}$ and zero rotation.  A schematic diagram of the wind configuration is shown in Figure \ref{fig:schematic} for the case $R_{\rm Y} = R_{\rm L}$.

Although our main goal is to assess magnetized PNS winds as an $r$-process site, we do not perform a detailed calculation of the nucleosynthetic yields here.  Rather, we aim to ascertain those dynamical and thermodynamical characteristics of the wind which are most relevant in setting the maximum atomic mass number to which the nuclear flow can proceed.  As already mentioned, the latter depends most sensitively on the ratio of neutrons to seed nuclei, which itself depends on the entropy $S$, electron fraction $Y_e$, and expansion timescale $t_{\rm exp}$ at radii where alpha particle formation completes \citep{Hoffman+97}.  

The wind dynamics is calculated under the (Newtonian) approximation of non-relativistic velocities, even though this assumption fails near the light cylinder.  This inaccuracy is tolerable because our concern is with the wind properties interior to where $T = 0.5$ MeV, as typically occurs within $\lesssim$ 10 PNS radii above the surface where special relativistic effects play no appreciable role.  For simplicity we also neglect general relativistic (GR) effects, except as included in an approximate way in a few test cases.  Past studies have shown that GR tends to increase the entropy of the flow (\citealt{Cardall&Fuller97}; \citealt{Otsuki+00}), but this effect is expected to be similar for rotating and non-rotating stars, and our main concern here is a comparison between these cases.

\subsection{Wind Equations \label{sec:windeq}}

The steady-state axisymmetric wind equations in cylindrical coordinates ($x$, $z$) are given by
\be
 \frac{\p}{\p l}\left(\frac{\rho v}{B}\right)=0 ;
\label{eq:mass}
\ee
\be
 -v\frac{\p v}{\p l}+x\Omega^2(\hat l \cdot \hat x) -\frac{1}{\rho}\frac{\p p}{\p l}-\frac{GM}{x^2+z^2}(\hat l\cdot \hat r)=0;
\label{eq:momentum}
\ee
\be
 v\frac{dY_e}{dl}=\lambda(\rho,T,Y_e) ;
\label{eq:ye}
\ee
\be
 vT\frac{dS}{dl}=\dot{q}(\rho,T,Y_e),
\label{eq:entropy}
\ee 
where $l$ is the distance measured along the field line.  Equation (\ref{eq:mass}) expresses mass continuity, where $\rho$ is the gas density, $v$ is the velocity along the magnetic field, $B = \sqrt{B_p^2+B_\phi^2}$ is the magnetic field strength (where $B_{\phi}$ and $B_p$ are the toroidal and poloidal components, respectively).  In $\S\ref{sec:magnmodel}$ we describe how the field strength and outflow geometry are calculated.  Equation (\ref{eq:momentum}) expresses momentum conservation, where the second term is the centrifugal force\footnote{The Coriolis force does not affect the motion of gas because it is perpendicular to the velocity and hence to the magnetic field, the latter of which is parallel to the flow velocity in the force-free approximation.}, $\Omega = 2\pi/P$ is the rotation rate of the PNS, $\hat l$ is the unit vector along the field line, and $\hat r$ is the unit vector from the center of the PNS (see Fig.~\ref{fig:schematic}).  The pressure $p$, entropy $S$ and other thermodynamic quantities are calculated from the Helmholtz equation of state (\citealt{Timmes&Swesty99}).

Equation (\ref{eq:ye}) evolves the electron fraction $Y_e$ due to weak interactions, which occur at the rate $\lambda(\rho,T,Y_e)$.  Equation (\ref{eq:entropy}) evolves the entropy of the outflowing gas due to neutrino heating and cooling, where $\dot{q}(\rho,T,Y_e)$ is the net heating rate.  Tabulated values of $\lambda$ and $\dot{q}$ are calculated as a function of density, temperature, neutrino luminosity $L_{\nu}$, and neutrinosphere temperature $T_{\nu}$  ($\S\ref{sec:nuphys}$).
 
Equation (\ref{eq:mass}) can be manipulated to read
\be
\frac{\p\rho}{\p l}=-\frac{\rho}{v}\frac{\p v}{\p r}+\rho\frac{d \ln(B)}{dl}, 
\label{eq:drhodl}
\ee
which upon substitution into equation (\ref{eq:momentum}) gives
\be
 \frac{c_s^2-v^2}{v}\frac{dv}{dl}=c_s^2\frac{d\ln(B)}{dl}+\frac{1}{\rho}\pt p l \rho+\frac{GM}{r^2+x^2}(\hat l \cdot \hat r) -r\Omega^2(\hat l \cdot \hat x) \label{veq}
\ee
where $c_s^2\equiv\pt p \rho {S,Y_e}$ and the second term on the right hand side can be written in terms of the other thermodynamic variables as follows
\begin{eqnarray}
 &&\pt p l \rho = \nonumber \\
&&\pt p S {\rho,Y_e}\frac{dS}{dl}+\pt p {Y_e} {\rho,S}\frac{dY_e}{dl}=v\left\{\frac{\dot{q}}{T}\pt p S {\rho,Y_e}+\lambda\pt p {Y_e} {\rho,S}\right\}, \nonumber \\
\end{eqnarray}
where we have used equations (\ref{eq:ye}) and (\ref{eq:entropy}).  

\subsection{Neutrino microphysics \label{sec:nuphys}}
Neutrino microphysics enters the wind evolution through the functions $\lambda$ and $\dot{q}$ (eqs.~[\ref{eq:ye}],[\ref{eq:entropy}]).  Included are the dominant charged current reactions: $\bar{\nu}_{\rm e}+p\rightarrow n+e^+,{\nu}_{\rm e}+n\rightarrow p+e, p+e\rightarrow n+\nu_{\rm e},$ and $n+e^+\rightarrow p+\bar{\nu}_{\rm e}$.  Cooling also includes neutral-current interactions
 (\citealt{Itoh+96}).  We neglect the effects of strong magnetic fields on the rate of neutrino heating and cooling in the PNS atmosphere (\citealt{Duan&Qian04}; \citealt{Riquelme+05}), as the corrections (primarily due to quantization of the lepton phase space into Landau levels) are found to be relatively minor for surface field strengths $\lesssim 10^{16}$ G.  The temperature of the neutrinosphere is calculated from the total electron neutrino luminosity according to 
\be
 L_{\nu_{e}}+L_{\bar\nu_e}= 4\pi R_{\rm ns}^{2}\frac{7\sigma}{8}T_{\nu}^{4} \label{eq:Tnu}.
\ee

Competition between the absorption of neutrinos and antineutrinos by nucleons in the wind (as dominate reactions well above the PNS surface) drive the electron fraction towards the equilibrium value 
\be
 Y_e^{\rm eq}=\left(1+\frac{L_{\bar{\nu}_e}}{L_{\nu_e}}\frac{\langle\eps_{\bar{\nu}_e}\rangle-2\Delta+1.2\Delta^2/\langle\eps_{\bar{\nu}_e}\rangle}{\langle\eps_{\nu_e}\rangle+2\Delta+1.2\Delta^2/\langle\eps_{\nu_e}\rangle}\right)^{-1},
\label{eq:Yeeq}
\ee
set by the luminosities $L_{\nu_{e}}$/$L_{\bar{\nu}_{e}}$ and mean energies $\langle\eps_{\bar{\nu}_e}\rangle/\langle\eps_{\nu_e}\rangle$ of the electron neutrinos and antineutrinos, respectively, where $\Delta=1.293$ MeV is the proton-neutron mass difference (\citealt{Qian&Woosley96}).  


Nucleosynthesis is extremely sensitive to the asymptotic value of $Y_e $, which in most of our wind solutions is found to approach $Y_e^{\rm eq}$ to high accuracy.  The value and temporal evolution of $Y_e^{\rm eq}$, however depends on theoretical uncertainties in the neutrino spectrum of the cooling PNS (e.g.~\citealt{MartinezPinedo+12}; \citealt{Roberts+12}).   Our calculations take the mean neutrino and antineutrino energies to be equal, $\langle\eps_{\bar{\nu}_e}\rangle=\langle\eps_{\nu_e}\rangle$, at the common value determined from the neutrinosphere temperature (eq.~[\ref{eq:Tnu}]).  The ratio of $L_{\nu_e}$ and $L_{\bar{\nu_e}}$ is then chosen such that $Y_{e}^{\rm eq} \simeq 0.45$ (eq.~[\ref{eq:Yeeq}]), as motivated by recent calculations which find a value of this order at early times in the PNS cooling evolution ($t \lesssim 5-10$ s after core bounce; e.g.~\citealt{Roberts+12}, their Fig.~5).

\subsection{Magnetosphere model \label{sec:magnmodel}}

\begin{figure}
\begin{center}
\includegraphics[width=1.0\linewidth]{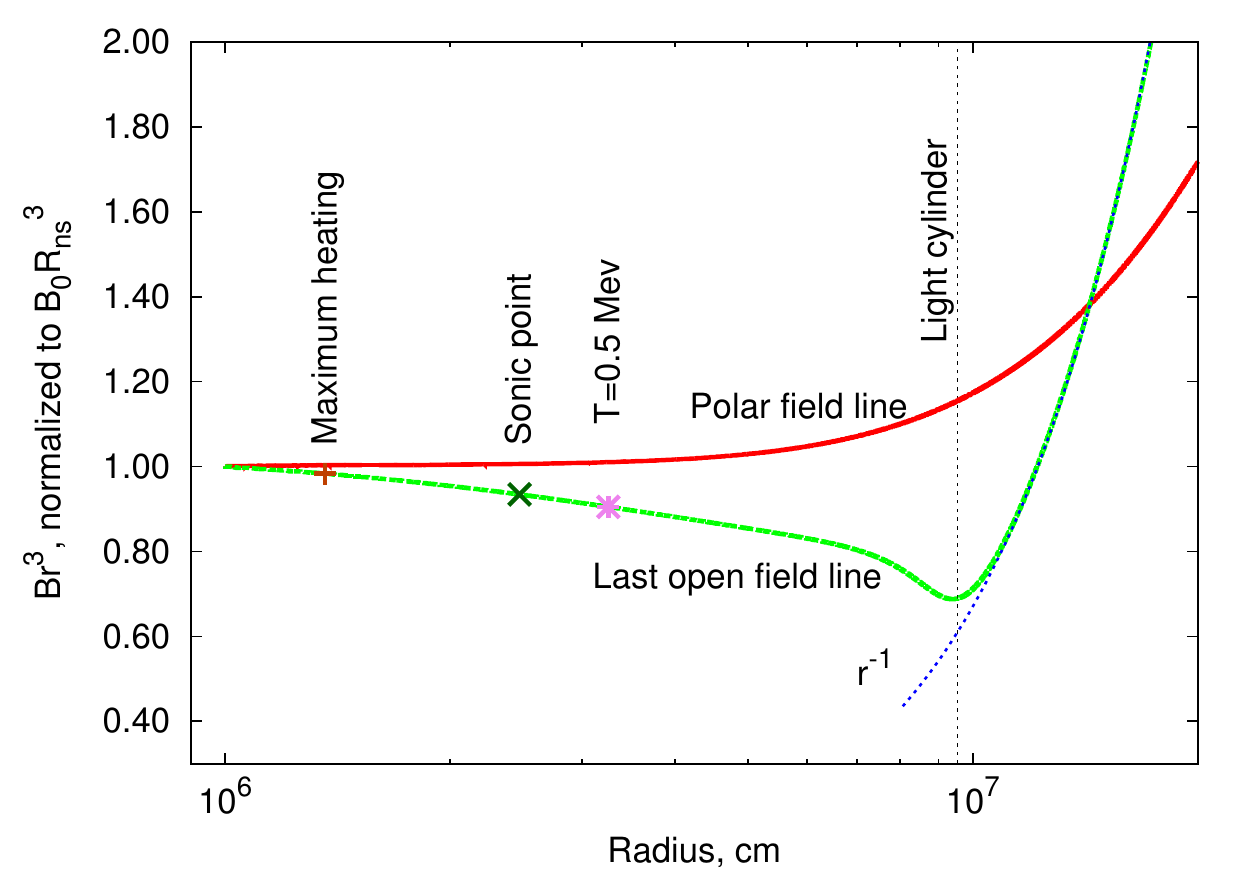}
\caption{
Magnetic field strength $B$ normalized to its surface value $B_{0}$, along two field lines, corresponding to a polar outflow ($\theta = 0$) and the last open field line ($\theta = \theta_{\rm max} \approx 21^{\circ}$), calculated for a PNS with spin period $P= 2$ ms.  Near the surface both lines obey $B \propto 1/r^3$ (dipole), while at larger radii the polar and last open field line follow $B \propto1/r^{2}$ and $\propto 1/r$, respectfully.  Shown for comparison are the locations of the sonic point ({\it cross}), maximum net heating ({\it plus}), alpha particle formation radius $R_{\rm 0.5MeV}$ ({\it asterisk}), and light cylinder radius $R_{\rm L}$ for our solution with $L_{\nu} = 10^{52}$ erg s$^{-1}$.
}
\label{fig:Br}
\end{center}
\end{figure}

The magnetic field strength $B$ and the field line trajectories (x,z) are calculated as a function of the distance along the field line $l$ according to (e.g.~\citealt{Timokhin06})
\begin{eqnarray}
  B(l)&=&\xi/x \\
 \left(\frac{dx}{dl},\frac{dz}{dl}\right)&=&\left(\hat l \cdot \hat x, \hat l \cdot \hat z\right)=\xi^{-1}\left(-\p_z\Psi,\p_x\Psi\right)\\
 \hat l\cdot\hat r &=& \frac{z\p_x\Psi-x\p_z\Psi}{\xi \sqrt{x^2+z^2}},
\ee
where 
\be \xi \equiv \sqrt{(\nabla\Psi)^2+\left(\frac{4\pi}{c}\right)^2I^2},
\ee
and $\Psi(x,z)$, $I(x,z)$ are functions proportional to the magnetic flux and current, respectively, flowing through a ring of constant cylindrical coordinates $(x,z)$. Coordinates $(x,z)$, quantities $l, \hat l, r, \hat r$ are illustrated at Figure \ref{fig:schematic}.  We employ the solutions\footnote{Under force-free conditions the current can be written as a function of the flux, such that $I(\Psi)$.} $\Psi(x,z)$ and $I(\Psi)$ of \citet{Timokhin06} for the standard case in which the last closed field line ($\theta = \theta_{\rm max}$) crosses the equatorial plane at the light cylinder ($R_{\rm Y} = R_{\rm L}$).  As will be discussed further in $\S\ref{sec:forcefree}$, the open zone may be larger than this ($R_{\rm Y} < R_{\rm L}$) due to the high pressure of the closed zone.

The geometry of the force-free magnetosphere is invariant to an overall rescaling of $\Psi$ and $I$.  This implies that our solutions do not depend on the strength of the magnetic field, as long as it is sufficiently high to justify the force-free condition ($\S\ref{sec:forcefree}$).  In contrast, our results do depend on the PNS rotation period, as the latter determines the physical radius of the light cylinder.

Figure \ref{fig:Br} shows the magnetic field strength $B$ as a function of spherical radius $r$, calculated for a PNS spin period $P= 2$ ms along two different field lines, corresponding to a polar outflow ($\theta = 0$) and the last open field line ($\theta = \theta_{\rm max} \approx 21^{\circ}$).  Close to the PNS surface, the poloidal component of the field is dipolar ($B_{\rm p} \propto 1/r^3$), with the toroidal component $B_{\phi}$ being small in comparison.  At larger radii near the light cylinder and beyond, the poloidal field becomes monopolar ($B_{\rm p} \propto 1/r^2$), while the toroidal field approaches the dependence $B_{\phi} \propto 1/r$ expected given the magneto-statics analogy between the polar current and an infinite wire.  These expectations are borne out in Figure \ref{fig:Br} by the magnetic field strength along the last open field line, which varies as $\propto 1/r^{3}$ close to the PNS surface (poloidal dominated) before transitioning to $\propto 1/r$ (toroidal dominated) near the light cylinder.  The polar field line $\theta = 0$, by contrast, is dominated by the poloidal component at all radii, which slowly transitions from dipolar $\propto 1/r^{3}$ to monopolar $\propto 1/r^{2}$.  

The faster the magnetic field strength decreases, the faster the matter expands. The magnetic field near the surface decreases faster in the proto-magnetar case $B\propto 1/r^3$ than for a spherical outflow $B\propto 1/r^2$. Non-polar outflows also experience additional centrifugal acceleration, which causes the matter to expand even more rapidly along field lines with larger $\theta$ ($\S$\ref{sec:results}). In $\S$\ref{sec:results} we show that centrifugal effects dominate the purely geometric effect of the diverging field geometry in terms of its influence on the conditions for a successful $r$-process.

The necessity of employing a realistic description for the field geometry is made clear by the locations of various radii in the flow  which are critical to the nucleosynthesis.  As shown by symbols in Figure \ref{fig:Br}, the sonic point, location of maximum heating, and alpha particle formation radius often occur at radii where the field geometry is not well-described by any limiting case.  

\subsection{Conservation checks}
\label{sec:conservation}
Equation (\ref{eq:mass}) shows that the ratio of mass to magnetic flux,
\be
 \dot{\mathcal{M}} \equiv \frac{\rho v}{B},
\label{eq:mdot}
\ee
is a conserved quantity and hence is automatically conserved at the level of our numerical scheme.  Conservation of momentum, energy, and lepton fraction, on the other hand, serve as independent checks of our solutions.  Momentum and energy conservation (eqs.~[\ref{eq:momentum}],[\ref{eq:entropy}]) can be combined into the integral constraint  (`Bernoulli integral') that 
\be
 \frac{\rho v}{B}\left(\frac{v^2}{2}+\eps+\frac{p}{\rho}-\frac{GM}{\sqrt{r^2+z^2}}-\frac12 r^2\Omega^2 \right)-\int_{l_0}^l\frac{\rho(l')\dot q(l')}{B(l')}dl',
\label{eq:bernoulli}
\ee 
is constant at each distance $l$ across the radial grid, where $\eps$ is the internal energy and $l_0$ is an arbitrary starting point.

\subsection{Shooting method} 

Equations (\ref{eq:ye}), (\ref{eq:entropy}), and (\ref{veq}) are solved using a `shooting' method by which the eigenvalue is iteratively guessed and corrected in order to converge upon an accurate solution.  Equation (\ref{veq}) has a singularity at the sonic point ($v=c_s$), which can only be avoided for a transonic solution if the right hand side of the equation simultaneously goes to zero.  The value of $\dot{\mathcal{M}}$ that zeroes the right hand side at $v=c_s$ is not known {\it a priori} but must instead be determined simultaneous with the full solution.  

Several considerations are used to determine whether a given guess for $\dot{\mathcal{M}}$ is bigger or smaller than the critical value.  First, if $dv/dl$ is negative at any point along the integration, then this implies that $\dot{\mathcal{M}}$ is smaller than the transonic value.  On the other hand, if $dv/dl$ exceeds an order-of-magnitude estimate of its value by a large factor (taken to be $\sim$ 50) then $dv/dl$ is assumed to be approaching infinity, indicating that $\dot{\mathcal{M}}$ is bigger than the transonic value.  Such a `guess and check' technique is used iteratively to determine $\dot{\mathcal{M}}$ and the structure of the subsonic solution up to the point where $v$ is a few percent less than $c_s$.  Above this velocity $dv/dl$ is taken to be constant in integrating the other equations (\ref{eq:ye},\ref{eq:entropy}) across the sonic point.  Finally, once $v$ is a few percent above $c_s$, then the full set of equations (\ref{eq:ye},\ref{eq:entropy},\ref{veq}) are again used to determine the supersonic structure.  Our solutions are found to conserve energy/momentum (eq.~[\ref{eq:bernoulli}]) to an accuracy of $\lesssim 3\%$.

In addition to the transonic condition, boundary conditions on the temperature, electron fraction, and density at the PNS surface are calculated according to the following criteria (\citealt{Metzger+07}):
\begin{enumerate} 
 \item Zero net heating ($\dot{q} = 0$.)
 \item Weak equilibrium ($dY_e/dt \simeq \lambda = 0$).
 \item Density $\rho = 10^{12}$ g cm$^{-3}$, as found to produce an integrated neutrino optical depth from infinity $\tau_{\nu} \sim 1$ (thus definining the surface as the `neutrinosphere').
\end{enumerate} 


\section{Results}
\label{sec:results}

Solutions are calculated for a range of total neutrino luminosities $L_{\nu} \equiv (L_{\nu_e} + L_{\bar{\nu_e}})/2 = 10^{52}, 6\times 10^{51}, 3\times 10^{51}, 10^{51}$ erg s$^{-1}$, as are physically achieved at times $t \sim 1$, $3$, $6$ and $15$ seconds after core bounce according to the PNS cooling calculations of \citet{Pons+99}.  For each luminosity, solutions are calculated for different values of the PNS spin period $P = 1, 2, 3, 5, 10$ ms and for different angles $\theta$ of the open field lines as a fraction of the angle $\theta_{\rm max}$ of the last open line (Fig.~\ref{fig:schematic}).

All results discussed in this section, as summarized in Figures \ref{fig:Bmin}$-$\ref{fig:mdotint} and Table \ref{table:summary}, are calculated assuming a PNS of `standard' mass $M=1.4 M_{\odot}$ and radius $R_{\rm ns}=12$ km (except where otherwise noted).  Results for a more massive PNS ($M=2 M_{\odot}$) are given in Tables \ref{table:2Msun} and \ref{table:integrated2Msun} in Appendix \ref{sec:appendix}.

Because only a fraction of the PNS surface is open to outflows in the magnetized rotating case, a {\it spherically equivalent} mass loss rate can be defined according to
\be
\dot{M} \equiv 4\pi R_{\rm ns}^{2}\rho_0 v_0,
\label{eq:Mdot}
\ee
where $\rho_0$ and $v_0$ are the density and velocity, respectively, at the PNS surface.  The value of $\dot{M}$ for each field line solution is reported in Tables \ref{table:summary},\ref{table:2Msun} and Figure \ref{fig:mdottwoms}.  The spherically equivalent mass loss rate $\dot{M}$ is not to be confused with the true total mass loss rate  
\be
 \dot{M}_{\rm tot}=\int \rho_0 v_0 dS,
\label{eq:mdottot}
\ee 
where $\rho_0, v_0$ are surface density and velocity, and the integration extends over the PNS surface threaded by open magnetic flux, $\theta \in [0,\theta_{\rm max}]$.  The values of $\dot{M}_{\rm tot}$ for each neutrino luminosity and spin period is reported in Tables \ref{table:integrated}, \ref{table:integrated2Msun} and Figure \ref{fig:mdotint}.

\begin{table*}
\begin{scriptsize}
\begin{center}
\vspace{0.05 in}\caption{Summary of Wind Models for $R_{NS}=12$ km, $M=1.4 M_{\odot}$}
\label{table:summary}
\begin{tabular}{cccccccccc}
\hline \hline
\multicolumn{1}{c}{$P$} &
\multicolumn{1}{c}{$L_{\nu}^{(a)}$} &
\multicolumn{1}{c}{$\frac{\theta}{\theta_{\rm max}}$} & 
\multicolumn{1}{c}{$\dot{M}^{(b)}$} & 
\multicolumn{1}{c}{$S^{(c)}$} &
\multicolumn{1}{c}{$t_{\rm exp}^{(d)}$} &
\multicolumn{1}{c}{$Y_e^{(e)}$} &
\multicolumn{1}{c}{$\eta = S^3/t_{\rm exp} Y_e^3$} &
\multicolumn{1}{c}{$R_{0.5\rm MeV}^{(f)}$} & 
\multicolumn{1}{c}{$\frac{P_{\rm gas}}{P_{\rm mag}}|_{R_{0.5\rm MeV}}^{(g)}$}\\
\hline
 (ms) & (10$^{51}$ erg s$^{-1}$)  &  - & ($M_{\odot}$ s$^{-1}$) & (k$_{\rm b}$ n$^{-1}$) & (ms) & - & $\%$ of ($\eta_{\rm thr}^{(h)}$) & (km) & -\\
\hline 
\\
$^{*}$N/A & 1 & $^{*}$N/A & $3.8\cdot 10^{-6}$ & 100 & $183.7$ & 0.4539 & 0.7 & 47 & $^{*}$N/A \\
$^{*}$N/A & 3 & $^{*}$N/A & $7.1\cdot 10^{-5}$ & 77 & $42.4$ & 0.4528 & 1.5 & 68 & $^{*}$N/A \\
$^{*}$N/A & 6 & $^{*}$N/A & $4.5\cdot 10^{-4}$ & 67 & $19.9$ & 0.4520 & 2.1 & 89 & $^{*}$N/A \\
$^{*}$N/A & 10 & $^{*}$N/A & $1.7\cdot 10^{-3}$ & 61 & $13.0$ & 0.4513 & 2.4 & 114 & $^{*}$N/A \\
$2^\dagger$ & 10 & 0.7 & $7.7\cdot 10^{-4}$ & 67 & 1.0 & 0.4519 & 40.6 & 35 & $3.3\cdot 10^{-2}$\\

2 & 3 & $0.0$ & $1.1\cdot 10^{-4}$ & 74 & 26.2 & 0.4525 & $2.2$ & 48 & $1.0\cdot 10^{-1}$\\
2 & 6 & $0.0$ & $7.4\cdot 10^{-4}$ & 65 & 10.9 & 0.4517 & $3.5$ & 58 & $3.3\cdot 10^{-1}$\\
2 & 10 & $0.0$ & $2.9\cdot 10^{-3}$ & 60 & 6.6 & 0.4508 & $4.6$ & 70 & $9.5\cdot 10^{-1}$\\
2 & 6 & $0.3$ & $9.3\cdot 10^{-4}$ & 48 & 2.3 & 0.4497 & $6.8$ & 41 & $6.6\cdot 10^{-2}$\\
2 & 10 & $0.3$ & $3.6\cdot 10^{-3}$ & 45 & 1.7 & 0.4483 & $7.9$ & 49 & $2.4\cdot 10^{-1}$\\
2 & 1 & $0.7$ & $9.6\cdot 10^{-6}$ & 55 & 16.6 & 0.4525 & $1.4$ & 25 & $3.1\cdot 10^{-3}$\\
2 & 3 & $0.7$ & $1.9\cdot 10^{-4}$ & 44 & 2.6 & 0.4492 & $4.7$ & 30 & $1.1\cdot 10^{-2}$\\
2 & 6 & $0.7$ & $1.2\cdot 10^{-3}$ & 40 & 1.3 & 0.4466 & $6.9$ & 35 & $4.5\cdot 10^{-2}$\\
2 & 10 & $0.7$ & $4.7\cdot 10^{-3}$ & 37 & 0.8 & 0.4447 & $9.5$ & 42 & $2.1\cdot 10^{-1}$\\
2 & 3 & $1.0$ & $2.5\cdot 10^{-4}$ & 37 & 1.6 & 0.4459 & $4.8$ & 27 & $9.7\cdot 10^{-3}$\\
2 & 10 & $1.0$ & $6.2\cdot 10^{-3}$ & 33 & 0.6 & 0.4398 & $9.0$ & 38 & $2.7\cdot 10^{-1}$\\
2 & 1 & $1.0$ & $1.3\cdot 10^{-5}$ & 45 & 9.2 & 0.4506 & $1.4$ & 23 & $2.4\cdot 10^{-3}$\\
2 & 3 & $1.0$ & $2.6\cdot 10^{-4}$ & 37 & 1.5 & 0.4454 & $4.8$ & 27 & $9.7\cdot 10^{-3}$\\
2 & 6 & $1.0$ & $1.7\cdot 10^{-3}$ & 33 & 0.8 & 0.4415 & $7.4$ & 32 & $4.7\cdot 10^{-2}$\\
2 & 10 & $1.0$ & $6.5\cdot 10^{-3}$ & 32 & 0.6 & 0.4390 & $9.1$ & 38 & $2.8\cdot 10^{-1}$\\
3 & 3 & $0.1$ & $1.1\cdot 10^{-4}$ & 75 & 26.7 & 0.4525 & $2.2$ & 48 & $4.7\cdot 10^{-2}$\\
3 & 6 & $0.1$ & $7.4\cdot 10^{-4}$ & 66 & 11.0 & 0.4517 & $3.6$ & 58 & $1.5\cdot 10^{-1}$\\
3 & 10 & $0.1$ & $2.9\cdot 10^{-3}$ & 61 & 6.5 & 0.4509 & $4.8$ & 69 & $4.4\cdot 10^{-1}$\\
3 & 1 & $0.3$ & $6.7\cdot 10^{-6}$ & 77 & 56.2 & 0.4536 & $1.1$ & 31 & $3.9\cdot 10^{-3}$\\
3 & 3 & $0.3$ & $1.3\cdot 10^{-4}$ & 61 & 9.3 & 0.4519 & $3.3$ & 39 & $1.5\cdot 10^{-2}$\\
3 & 6 & $0.3$ & $8.4\cdot 10^{-4}$ & 54 & 3.8 & 0.4507 & $5.8$ & 46 & $4.7\cdot 10^{-2}$\\
3 & 10 & $0.3$ & $3.2\cdot 10^{-3}$ & 51 & 2.7 & 0.4499 & $7.2$ & 56 & $1.7\cdot 10^{-1}$\\
3 & 1 & $0.7$ & $7.7\cdot 10^{-6}$ & 66 & 31.8 & 0.4533 & $1.3$ & 28 & $2.3\cdot 10^{-3}$\\
3 & 3 & $0.7$ & $1.5\cdot 10^{-4}$ & 53 & 5.0 & 0.4511 & $4.1$ & 34 & $8.5\cdot 10^{-3}$\\
3 & 10 & $0.7$ & $3.8\cdot 10^{-3}$ & 44 & 1.4 & 0.4480 & $8.6$ & 47 & $1.1\cdot 10^{-1}$\\
3 & 1 & $1.0$ & $9.2\cdot 10^{-6}$ & 58 & 19.8 & 0.4528 & $1.4$ & 26 & $1.7\cdot 10^{-3}$\\
3 & 6 & $1.0$ & $1.2\cdot 10^{-3}$ & 42 & 1.4 & 0.4475 & $7.6$ & 36 & $2.4\cdot 10^{-2}$\\
3 & 10 & $1.0$ & $4.5\cdot 10^{-3}$ & 39 & 0.9 & 0.4457 & $9.5$ & 43 & $1.1\cdot 10^{-1}$\\
4 & 3 & $0.1$ & $1.1\cdot 10^{-4}$ & 75 & 26.9 & 0.4524 & $2.2$ & 48 & $2.7\cdot 10^{-2}$\\
4 & 6 & $0.1$ & $7.3\cdot 10^{-4}$ & 66 & 11.1 & 0.4517 & $3.6$ & 58 & $8.8\cdot 10^{-2}$\\
4 & 10 & $0.1$ & $2.9\cdot 10^{-3}$ & 61 & 6.5 & 0.4509 & $4.8$ & 69 & $2.5\cdot 10^{-1}$\\
4 & 1 & $0.4$ & $6.4\cdot 10^{-6}$ & 81 & 69.2 & 0.4537 & $1.0$ & 32 & $2.7\cdot 10^{-3}$\\
4 & 3 & $0.4$ & $1.2\cdot 10^{-4}$ & 64 & 11.9 & 0.4521 & $3.0$ & 41 & $1.1\cdot 10^{-2}$\\
4 & 6 & $0.4$ & $8.1\cdot 10^{-4}$ & 56 & 4.8 & 0.4510 & $5.2$ & 48 & $3.5\cdot 10^{-2}$\\
4 & 10 & $0.4$ & $3.2\cdot 10^{-3}$ & 52 & 2.9 & 0.4500 & $7.0$ & 57 & $1.1\cdot 10^{-1}$\\
4 & 3 & $0.7$ & $1.4\cdot 10^{-4}$ & 57 & 6.8 & 0.4516 & $3.7$ & 36 & $6.5\cdot 10^{-3}$\\
4 & 10 & $0.7$ & $3.5\cdot 10^{-3}$ & 47 & 1.8 & 0.4489 & $8.1$ & 49 & $6.2\cdot 10^{-2}$\\
4 & 1 & $1.0$ & $7.9\cdot 10^{-6}$ & 66 & 31.1 & 0.4533 & $1.3$ & 28 & $1.4\cdot 10^{-3}$\\
4 & 3 & $1.0$ & $1.5\cdot 10^{-4}$ & 52 & 4.9 & 0.4510 & $4.1$ & 34 & $5.0\cdot 10^{-3}$\\
4 & 6 & $1.0$ & $10.0\cdot 10^{-4}$ & 47 & 2.1 & 0.4493 & $7.1$ & 40 & $1.7\cdot 10^{-2}$\\
4 & 10 & $1.0$ & $3.9\cdot 10^{-3}$ & 44 & 1.4 & 0.4478 & $9.0$ & 47 & $6.7\cdot 10^{-2}$\\
5 & 6 & $0.1$ & $7.3\cdot 10^{-4}$ & 66 & 11.2 & 0.4517 & $3.6$ & 58 & $5.7\cdot 10^{-2}$\\
5 & 10 & $0.1$ & $2.9\cdot 10^{-3}$ & 61 & 6.5 & 0.4509 & $4.9$ & 69 & $1.6\cdot 10^{-1}$\\
5 & 6 & $0.4$ & $7.7\cdot 10^{-4}$ & 59 & 6.2 & 0.4513 & $4.7$ & 51 & $3.0\cdot 10^{-2}$\\
5 & 10 & $0.4$ & $3.0\cdot 10^{-3}$ & 55 & 3.7 & 0.4504 & $6.3$ & 61 & $8.9\cdot 10^{-2}$\\
5 & 1 & $0.6$ & $6.6\cdot 10^{-6}$ & 78 & 58.5 & 0.4536 & $1.1$ & 31 & $1.5\cdot 10^{-3}$\\
5 & 3 & $0.6$ & $1.3\cdot 10^{-4}$ & 61 & 9.8 & 0.4519 & $3.3$ & 39 & $6.0\cdot 10^{-3}$\\
5 & 6 & $0.6$ & $8.3\cdot 10^{-4}$ & 54 & 3.9 & 0.4508 & $5.7$ & 46 & $1.9\cdot 10^{-2}$\\
5 & 1 & $1.0$ & $7.4\cdot 10^{-6}$ & 71 & 41.1 & 0.4535 & $1.2$ & 29 & $1.1\cdot 10^{-3}$\\
5 & 3 & $1.0$ & $1.4\cdot 10^{-4}$ & 56 & 6.6 & 0.4515 & $3.8$ & 36 & $4.3\cdot 10^{-3}$\\
5 & 6 & $1.0$ & $9.3\cdot 10^{-4}$ & 50 & 2.7 & 0.4501 & $6.5$ & 42 & $1.4\cdot 10^{-2}$\\
5 & 10 & $1.0$ & $3.6\cdot 10^{-3}$ & 47 & 1.8 & 0.4487 & $8.3$ & 50 & $5.0\cdot 10^{-2}$\\
1 & 3 & $0.0$ & $1.1\cdot 10^{-4}$ & 72 & 23.3 & 0.4524 & $2.2$ & 48 & $3.2\cdot 10^{-1}$\\
1 & 6 & $0.0$ & $7.4\cdot 10^{-4}$ & 63 & 9.9 & 0.4515 & $3.5$ & 59 & $9.9\cdot 10^{-1}$\\
1 & 10 & $0.0$ & $2.9\cdot 10^{-3}$ & 57 & 6.2 & 0.4508 & $4.3$ & 72 & $2.8\cdot 10^{0}$\\
1 & 6 & $0.4$ & $1.7\cdot 10^{-3}$ & 32 & 0.7 & 0.4406 & $6.9$ & 32 & $1.6\cdot 10^{-1}$\\
1 & 1 & $0.7$ & $4.0\cdot 10^{-5}$ & 26 & 2.4 & 0.4312 & $1.2$ & 19 & $8.7\cdot 10^{-3}$\\
1 & 3 & $0.7$ & $8.7\cdot 10^{-4}$ & 21 & 0.6 & 0.4124 & $3.2$ & 22 & $4.9\cdot 10^{-2}$\\
1 & 10 & $0.7$ & $2.1\cdot 10^{-2}$ & 20 & 0.3 & 0.4001 & $5.3$ & 35 & $3.6\cdot 10^{0}$\\
1 & 3 & $1.0$ & $5.4\cdot 10^{-3}$ & 14 & 0.3 & 0.3191 & $4.1$ & 20 & $2.6\cdot 10^{-1}$\\
1 & 6 & $1.0$ & $3.2\cdot 10^{-2}$ & 14 & 0.2 & 0.3133 & $5.4$ & 25 & $2.8\cdot 10^{0}$\\
1 & 10 & $1.0$ & $1.1\cdot 10^{-1}$ & 14 & 0.2 & 0.3219 & $6.0$ & 32 & $2.7\cdot 10^{1}$\\

\\
\hline
\hline
\end{tabular}
\end{center}
$^{*}$Spherical solutions; $^{\dagger}$ Calculated using Pacynzsky-Wiita gravitational potential to approximate effects of GR.  $^{(a)}$ $L_{\nu} \equiv (L_{\nu_e} + L_{\bar{\nu}_e})/2$; $^{(b)}$ Mass outflow rate (spherical, or equivalent spherical; eq.~[\ref{eq:Mdot}]); $^{(c)}$Final entropy $^{(d)}$Expansion timescale at $T = 0.5$ MeV (eq.~[\ref{eq:texp}]); $^{(e)}$Final electron fraction; $^{(f)}$Radius at which $T = 0.5$ MeV, approximate location of alpha particle formation; $^{(g)}$Ratio of `matter pressure' $P_{\rm gas}$ to magnetic pressure $P_{\rm mag} = B^{2}/8\pi$ (for an assumed surface field $B_0 = 10^{15}$ G) at $T = 0.5$ MeV, where $P_{gas}$ includes gas pressure $P$, thermal energy density $e$, kinetic energy density (rest energy subtracted); $^{(h)}$Threshold ratio for outflow to attain third-peak $r$-process (eq.~[\ref{eq:eta}]);  
\end{scriptsize}
\end{table*}

\begin{table*}
\begin{scriptsize}
\begin{center}
\vspace{0.05 in}\caption{Summary of Integrated Quantities for $R_{NS}=12$ km, $M=1.4 M_{\odot}$}
\label{table:integrated}
\begin{tabular}{cccccc}
\hline \hline
\multicolumn{1}{c}{$P$} &
\multicolumn{1}{c}{$L_{\nu}^{(a)}$} &
\multicolumn{1}{c}{$\dot{M}_{\rm tot}^{(b)}$} & 
\multicolumn{1}{c}{$M_{\rm ej}^{(c)}$} &
\multicolumn{1}{c}{$\Gamma_{\rm max}^{(d)}$} &
\multicolumn{1}{c}{$\langle \eta \rangle = \langle S^{3}/t_{\rm exp}Y_{e}^{3}\rangle^{(e)}$} \\
\hline
 (ms) & (10$^{51}$ erg s$^{-1}$) &  ($M_{\odot}$ s$^{-1}$) & ($M_{\odot}$) & - & ($\%$ of $\eta_{\rm thr}$) \\
\hline 
\\\
$^{*}$N/A & 1 & $3.8\cdot 10^{-6}$ & $3.5\cdot 10^{-5}$ & $^{*}$N/A & 0.7 \\
$^{*}$N/A & 3 & $7.1\cdot 10^{-5}$ & $2.6\cdot 10^{-4}$ & $^{*}$N/A & 1.5 \\
$^{*}$N/A & 6 & $4.5\cdot 10^{-4}$ & $9.4\cdot 10^{-4}$ & $^{*}$N/A & 2.1 \\
$^{*}$N/A & 10 & $1.7\cdot 10^{-3}$ & $3.3\cdot 10^{-3}$ & $^{*}$N/A & 2.4 \\

1 & 1 & $5.5\cdot 10^{-6}$ & $5.1\cdot 10^{-5}$ & 40.4 & 1.2 \\
1 & 10 & $5.5\cdot 10^{-3}$ & $1.0\cdot 10^{-2}$ & 1.0 & 5.8 \\
1 & 3 & $2.7\cdot 10^{-4}$ & $9.8\cdot 10^{-4}$ & 1.8 & 3.9 \\
1 & 6 & $2.1\cdot 10^{-3}$ & $4.4\cdot 10^{-3}$ & 1.1 & 5.4 \\
2 & 1 & $6.8\cdot 10^{-7}$ & $6.3\cdot 10^{-6}$ & 20.9 & 1.4 \\
2 & 10 & $3.1\cdot 10^{-4}$ & $5.9\cdot 10^{-4}$ & 1.0 & 8.9 \\
2 & 3 & $1.3\cdot 10^{-5}$ & $4.6\cdot 10^{-5}$ & 2.1 & 4.4 \\
2 & 6 & $8.1\cdot 10^{-5}$ & $1.7\cdot 10^{-4}$ & 1.2 & 7.0 \\
3 & 1 & $3.8\cdot 10^{-7}$ & $3.5\cdot 10^{-6}$ & 8.0 & 1.3 \\
3 & 10 & $1.9\cdot 10^{-4}$ & $3.5\cdot 10^{-4}$ & 1.0 & 8.5 \\
3 & 3 & $7.0\cdot 10^{-6}$ & $2.6\cdot 10^{-5}$ & 1.4 & 3.9 \\
3 & 6 & $4.8\cdot 10^{-5}$ & $10.0\cdot 10^{-5}$ & 1.1 & 6.6 \\
4 & 1 & $2.8\cdot 10^{-7}$ & $2.6\cdot 10^{-6}$ & 4.0 & 1.1 \\
4 & 10 & $1.4\cdot 10^{-4}$ & $2.6\cdot 10^{-4}$ & 1.0 & 7.9 \\
4 & 3 & $5.5\cdot 10^{-6}$ & $2.0\cdot 10^{-5}$ & 1.2 & 3.6 \\
4 & 6 & $3.5\cdot 10^{-5}$ & $7.3\cdot 10^{-5}$ & 1.0 & 6.0 \\
5 & 1 & $2.4\cdot 10^{-7}$ & $2.2\cdot 10^{-6}$ & 2.4 & 1.1 \\
5 & 10 & $1.2\cdot 10^{-4}$ & $2.2\cdot 10^{-4}$ & 1.0 & 7.2 \\
5 & 3 & $4.7\cdot 10^{-6}$ & $1.8\cdot 10^{-5}$ & 1.1 & 3.4 \\
5 & 6 & $3.0\cdot 10^{-5}$ & $6.3\cdot 10^{-5}$ & 1.0 & 5.7 \\
10 & 10 & $4.4\cdot 10^{-5}$ & $8.3\cdot 10^{-5}$ & 1.0 & 5.7 \\
10 & 3 & $1.7\cdot 10^{-6}$ & $6.5\cdot 10^{-6}$ & 1.0 & 2.5 \\
10 & 6 & $1.1\cdot 10^{-5}$ & $2.4\cdot 10^{-5}$ & 1.0 & 4.2 \\

 \\
\hline
\hline
\end{tabular}
\end{center}
$^{*}$Spherical Solutions. 
$^{(a)}$ $L_{\nu} \equiv (L_{\nu_e} + L_{\bar{\nu}_e})/2$; $^{(b)}$Total mass loss rate, integrated across the open zone from $\theta = 0 - \theta_{\rm max}$ (eq.~[\ref{eq:mdottot}]).  $^{(c)}$Total ejected mass at luminosities $\sim L_{\nu}$, estimated using the PNS cooling evolution from \citet{Pons+99}.  $^{(d)}$Maximum Lorentz factor achieved by the proto-magnetar outflow, calculated assuming a surface field $B_{0} = 3\times 10^{15}$ G. $^{(e)}$Threshold for $r$-process (eq.~	[\ref{eq:eta}]), in a mass-weighted average over the open zone from $\theta = 0 - \theta_{\rm max}$.
\end{scriptsize}
\end{table*}


\subsection{Validity of the Force-Free Approximation}
\label{sec:forcefree}

Figure \ref{fig:Bcomp} compares the pressures associated with matter and radiation $P_{\rm gas}$ to that of the magnetic field $P_{\rm B} = B^{2}/8\pi$ as a function of radius, calculated for our fiducial solution ($L_{\nu} = 10^{52}$ erg s$^{-1}$; $P = 2$ ms) and an assumed surface magnetic field strength $B_0 = 10^{15}$ G.  Thermal pressure exceeds magnetic pressure at the PNS surface, as expected since otherwise the field could not be anchored to the star.  Gas pressure decreases rapidly with radius, such that magnetic pressure comes to dominate just $\sim 2-3$ km above the surface. Magnetic pressure continues to dominate until the kinetic energy grows to a comparable size near the light cylinder, at which point the force-free approximation breaks down (our assumption of non-relativistic velocity is violated at a similar location).  As already discussed, such inaccuracies are tolerated at large radii because we are interested is the outflow structure at temperatures $T \gtrsim 0.5$ MeV (marked as a circle in Fig.~\ref{fig:Bcomp}), at which point the value of $P_{\rm gas}/P_{\rm mag}$ is still $\lesssim 10$ per cent.  The ratio $P_{\rm gas}/P_{\rm mag}$ at $T=0.5$ MeV for each of our solutions is provided in Table \ref{table:summary}, for the same assumed surface field strength $B_0 = 10^{15}$ G.   

Figure \ref{fig:Bmin} shows the minimum surface field strength $B_{\rm min}$ for which the force-free criterion (defined as $P_{\rm gas}/P_{\rm mag} < 0.1$ at $T = 0.5$ MeV) is satisfied for all outflow angles $\theta < \theta_{\rm max}$, as a function of the neutrino luminosity and PNS spin period.  At the highest neutrino luminosities $\sim 10^{52}$ ergs, and hence earliest times, our solutions provide an accurate description of the outflow at radii of interest only for strong fields $B_{0} \gtrsim 3\times 10^{15}$ G (for $P \gtrsim 2$ ms).  However, at later times when $L_{\nu} \sim 10^{51}$ erg s$^{-1}$, even weaker fields $B_0 \sim  3\times 10^{14}$ G are sufficient to satisfy the force-free condition.

\begin{figure}
\begin{center}
\includegraphics[width=1.0\linewidth]{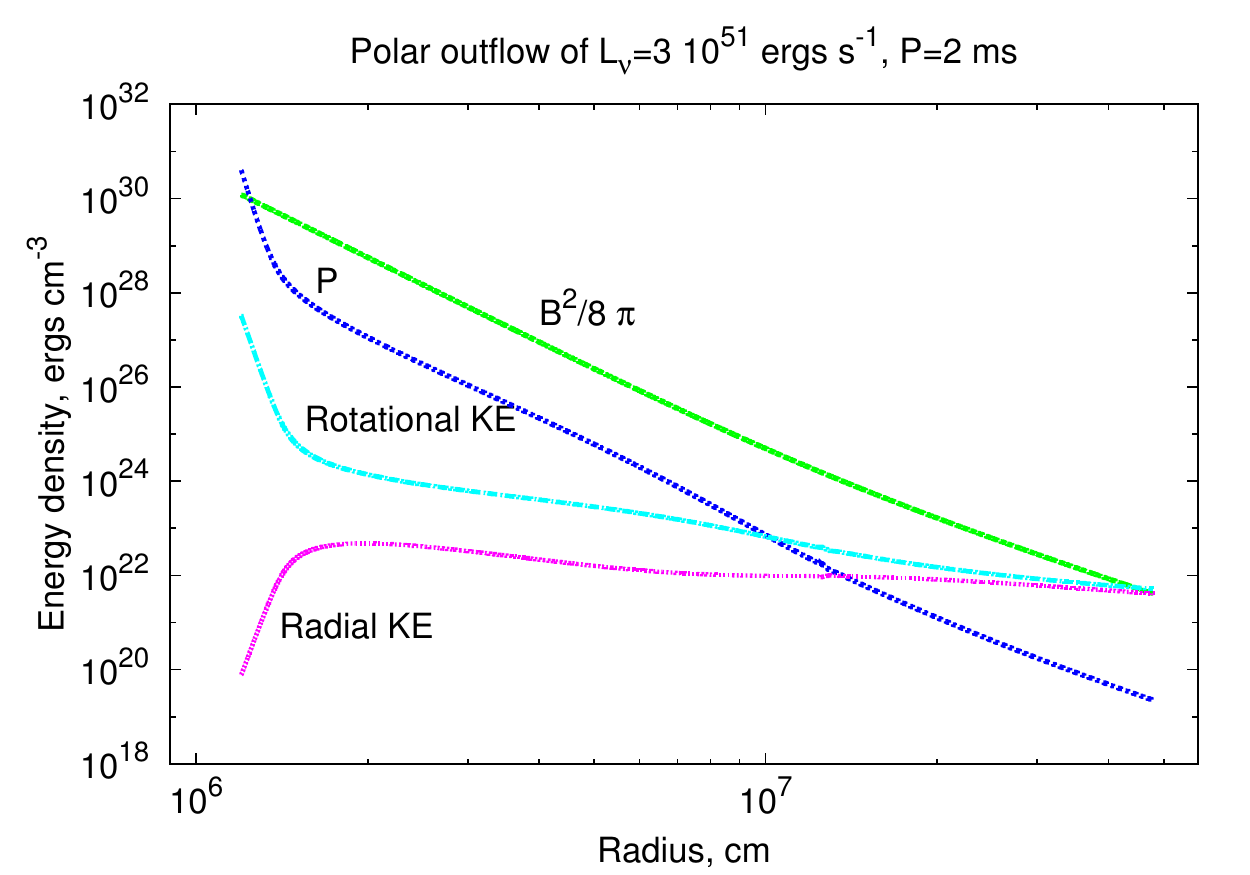}
\caption{Energy densities as a function of radius along the polar direction $\theta = 0$ for our fiducial solution with $L_{\nu} = 10^{52}$ erg s$^{-1}$ and $P = 2$ ms, assuming a surface field $B_{0} = 10^{15}$ G.
}
\label{fig:Bcomp}
\end{center}
\end{figure}

\begin{figure}
\begin{center}
\includegraphics[width=1.0\linewidth]{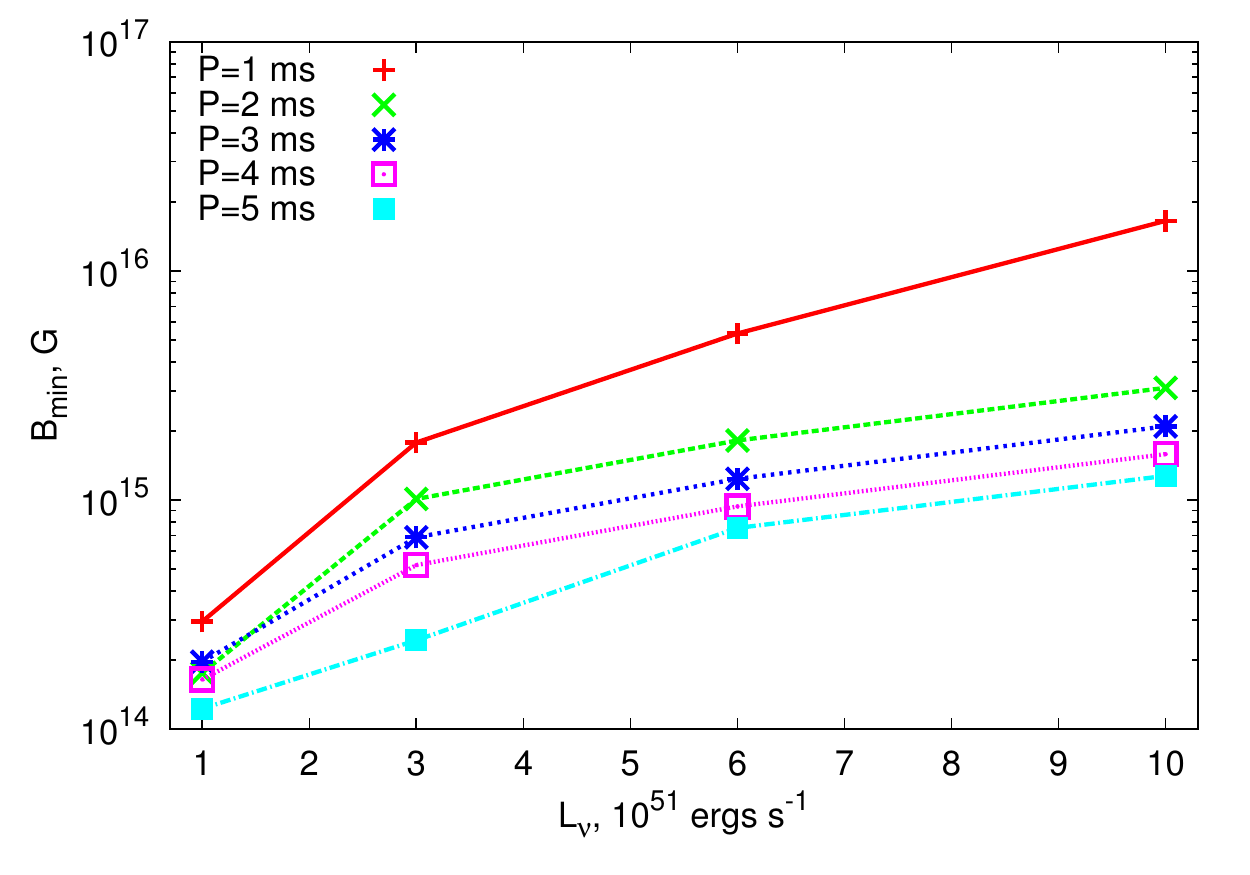}
\caption{Minimum surface magnetic field strength $B_{\rm min}$ required for validity of the force-free approximation at all angles and radii interior to the location of $\alpha$-particle formation ($T = 0.5$ MeV), calculated for different values of the neutrino luminosity $L_{\nu}$ and spin period $P$.  The force-free condition is defined as $P_{\rm gas}/P_{\rm mag}\le 0.1$, where $P_{\rm mag} = B^{2}/8\pi$ is the magnetic pressure and $P_{\rm gas}$ is the `matter pressure', the latter of which includes gas pressure $P$, thermal energy density $e$, and kinetic energy density (rest energy subtracted).}
\label{fig:Bmin}
\end{center}
\end{figure}

Even if the force-free approximation is valid in the outflow for magnetar-strength fields, the same may not be true in the closed zone due to the large {\it hydrostatic} pressure of the co-rotating atmosphere.  If the closed zone pressure $P_{\rm closed}$ exceeds that which can be confined by the magnetosphere, the field will `tear' open, enlarging the extent of the open zone ($R_{\rm Y} \lesssim R_{\rm L}$) compared to the standard force-free case $R_{\rm Y} = R_{\rm L}$ (\citealt{Thompson+04}; \citealt{Bucciantini+06}).  

Figure \ref{fig:hydrostatic} compares the radial profile of $P_{\rm closed}$ and $P_{\rm open}$ for $L_{\nu} = 3\times 10^{51}$ erg s$^{-1}$ and $P = 2$ ms along the last closed field line $\theta = \theta_{\rm max}$, to the pressure of the poloidal magnetic field $P_{\rm B} = B_{\rm p}^{2}/8\pi$, calculated for $B_0 = 10^{15}$ G.   The closed zone pressure is determined by solving the equations of hydrostatic, thermal and weak interactions equilibrium ($v = 0$; $\dot{q}=0; dY_e/dt=\lambda=0$).  The analytic model of \citet{Mestel&Spruit87} predicts that the closed zone $R_{\rm Y}$ extends to the location where $P_{\rm B} = P_{\rm closed}$, which according to Fig.~\ref{fig:hydrostatic} occurs at a radius $r \sim 30$ km for $B_0 = 10^{15}$ G which is a factor of $\sim 3$ times smaller than in the force-free limit ($R_{\rm Y} = R_{\rm L} = 100$ km).  Although most models in this paper assume $R_{\rm Y} = R_{\rm L} $, the true extent of the open zone $\theta_{\rm max}$ could be larger than our assumed value by a factor $(R_{\rm Y}/R_{\rm L})^{1/2} \lesssim 2$.  

\begin{figure}
\begin{center}
\includegraphics[width=1.0\linewidth]{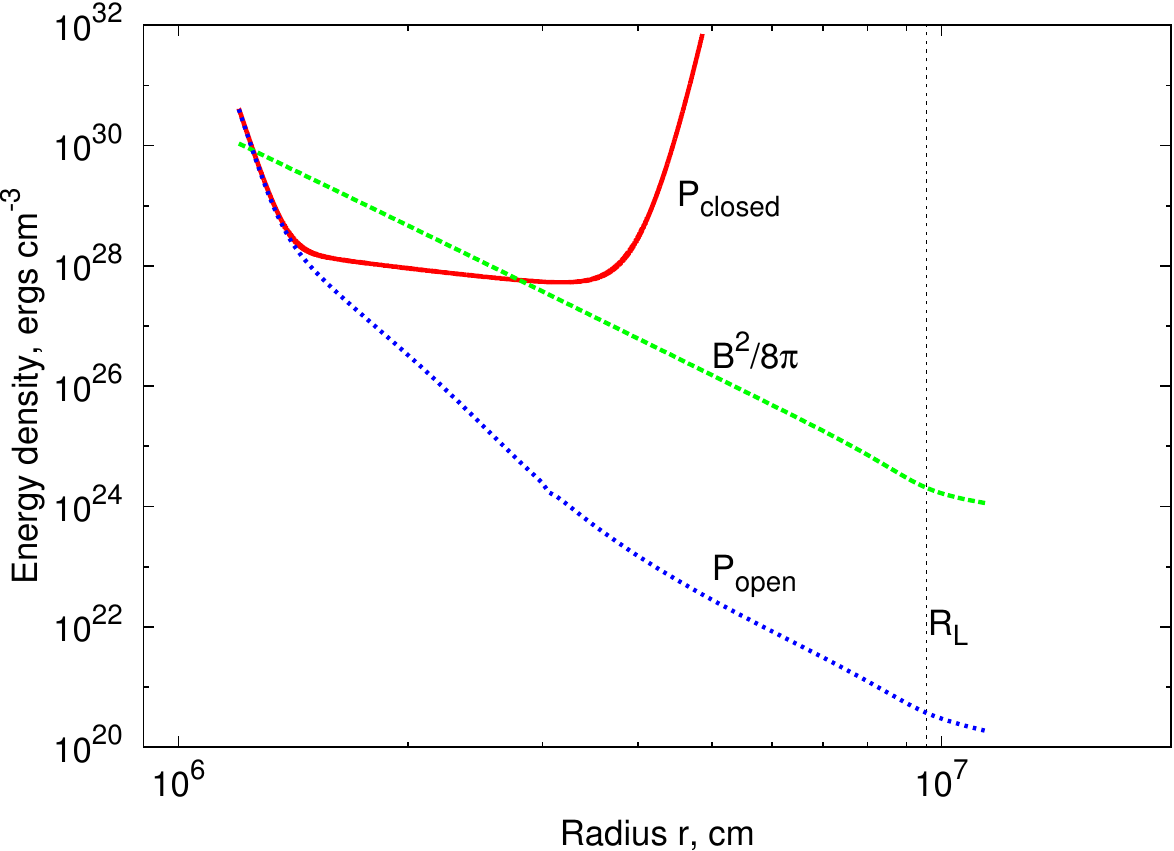}
\caption{Comparison between the hydrostatic thermal pressure of the closed zone $P_{\rm closed}$ ({\it red solid}) and the poloidal magnetic field pressure $B_{\rm p}^{2}/8\pi$ ({\it green dashed}) along the last open field line ($\theta = \theta_{\rm max}$), calculated for a PNS with neutrino luminosity $L_{\nu} = 3\times 10^{51}$ erg s$^{-1}$, spin period $P = 2 $ ms, and surface magnetic field strength $B_0 = 10^{15}$ G.  Also shown for comparison is the thermal pressure in the outflow itself $P_{\rm open}$ ({\it blue dotted}).  The closed zone pressure exceeds that of the confining magnetosphere at radii $\gtrsim 30$ km smaller than the light cylinder radius ($R_{\rm L} \approx 100$ km), which according to the criterion of \citep{Mestel&Spruit87} indicates that the true open portion of the magnetosphere may be larger than the force free assumption, i.e. the Y-point radius should obey $R_{\rm Y} < R_{\rm L}$ (see Fig.~\ref{fig:schematic}).}
\label{fig:hydrostatic}
\end{center}
\end{figure}

\subsection{Spherical versus Magnetized Rotating Outflows}

This section provides a comparison between the properties of our rotating PNS wind solutions (for fiducial parameters $L_{\nu} = 10^{52}$ erg s$^{-1}$, $P = 2$ ms) and the spherical non-rotating solutions corresponding to the same neutrino luminosity.  In particular, Figures  \ref{fig:multidyn} and \ref{fig:multitherm} show the radial profiles of wind properties along two field lines, corresponding to a polar outflow ($\theta = 0$) and one along the last open field line ($\theta = \theta_{\rm max} \approx 21^{\circ}$).  This section also compares the properties of our rotating solutions, such as the  entropy/expansion timescale (Fig.~\ref{fig:svstautwoms}) and mass loss rate (Fig.~\ref{fig:mdottwoms}), across a range of rotation periods and neutrino luminosities.   

\subsubsection{Final electron fraction}
\label{sec:Ye}

The electron fraction $Y_e$ quickly rises and saturates just a few kilometers of the PNS surface (Fig.~\ref{fig:multitherm}, top panel), with asymptotic values of $Y_e = 0.4523$ and $0.4504$ for the polar ($\theta = 0$) and inclined ($\theta = \theta_{\rm max}$) outflows, respectively, as compared to $Y_e = 0.4526$ for the otherwise equivalent spherical non-rotating case.  These values are nearly equal because in each case the outflow composition enters equilibrium with neutrino absorptions, which drives $Y_e$ to $Y_e^{\rm eq}$ (eq.~[\ref{eq:Yeeq}]).  Electron and positron captures play no significant role in the freeze-out process because their rate $\propto T^{6}$ decreases rapidly with temperature above the PNS surface, making them quickly negligible compared to neutrino absorptions.   

It is inevitable that the outflow composition enter equilibrium with the neutrinos if the latter dominate at freeze-out.  Each nucleon must absorb an energy $\gtrsim GM m_p/R_{\rm ns} \approx 160 $ MeV (for $M=1.4 M_{\odot}$ and $R_{\rm ns}=12$ km) to escape the gravitational potential well of the PNS.  Because this energy greatly exceeds that of an average neutrino  $\langle\eps_\nu\rangle \simeq 4kT_{\nu} \approx 15$ MeV, each outflowing nucleon must absorb $\sim 10$ neutrinos.  This process renders its initial identity as a proton or neutron forgotten (e.g.~\citealt{Metzger+08a}).  

Centrifugal acceleration alters the above argument by providing an additional source of wind energy independent of neutrino heating.  In practice, however, the asymptotic value of $Y_e$ is appreciably reduced below $Y_{e}^{\rm eq}$ only for extremely rapid rotation near the centrifugal break-up limit \citep{Metzger+08a}.  In our mostly rapidly rotating model with $P= 1$ ms, asymptotic $Y_e$ reaches values as low as $\sim 0.31-0.32$ for the largest outflow angles $\theta = \theta_{\rm max}$ (Table \ref{table:summary}).  However, for slower rotation $P \gtrsim 2$ ms, $Y_e$ varies by less than a percent from its spherical, non-rotating value across the entire open zone. 

\subsubsection{Mass loss rate}
\label{sec:mdot}
The spherically-equivalent mass loss rates of our fiducial polar ($\theta = 0$) and inclined ($\theta = \theta_{\rm max}$) solutions are $\dot{M} = 1\times 10^{-3}M_{\odot}$ s$^{-1}$ and $2\times 10^{-3} M_{\odot}$ s$^{-1}$, respectively, as compared to $\dot{M} = 7\times 10^{-4} M_{\odot}$ s$^{-1}$ for the spherical, non-rotating case.  These nearly equal values again result because the mass loss per unit surface area is set largely by the neutrino energy absorbed near the surface (\citealt{Qian&Woosley96}).  The structure of the outflow, being subsonic in the heating region, is determined mainly by hydrostatic equilibrium and hence is relatively insensitive to the precise outflow geometry.  

This expectation is borne out in the radial profiles of density (Fig.\ref{fig:multidyn}; {\it middle panel}) and heating (Fig.\ref{fig:multitherm}; {\it bottom panel}) just above the PNS surface, which are similar between the spherical and rotating magnetized outflows. The slightly larger value of $\dot{M}$ in the $\theta = \theta_{\rm max}$ case is the result of centrifugal acceleration, which increases the density scale-height in the net heating region, to which $\dot{M}$ is exponentially sensitive.  This increase is not as great as for a equatorial monopolar outflow with the same PNS rotational period (\citealt{Metzger+07}), because of the lower surface rotational velocity $v_{\phi,0} \lesssim R_{\rm ns}\Omega\sin \theta_{\rm max}$ as compared to the equatorial value $v_{\phi,0} = R_{\rm ns}\Omega$.

Figure \ref{fig:mdottwoms} compares the values of $\dot{M}$ for different rotation rates and neutrino luminosities. As expected, the mass loss rate depends sensitively on the neutrino luminosity, approximately as $\dot{M} \propto L_{\nu}^{5/3}\langle \epsilon_\nu\rangle^{10/3} \propto  L_{\nu}^{5/2}$, with a weaker dependence on other parameters (\citealt{Qian&Woosley96}).   The largest enhancement in $\dot{M}$ due to rotation is for the shortest rotation period ($P = 1$ ms) and largest angle ($\theta = \theta_{\rm max}$), for which $\dot{M}$ is a factor $\approx 10$ times larger than the polar or spherical cases.  Although the mass loss rate {\it per surface area} is enhanced by magnetic fields and rotation, this is offset by the smaller fraction of the PNS surface $f_{\rm open}$ which is open to outflows in the magnetized case: 
\be
f_{\rm open} \approx \frac{2 \pi \theta_{\rm max}^{2}}{4\pi} \approx \frac{1}{2}\frac{R_{\rm ns}}{R_{\rm Y}} \approx 0.13\frac{R_{\rm L}}{R_{\rm Y}}\left(\frac{P}{\rm ms}\right)^{-1},
\label{eq:fopen}
\ee 
where in the final two equalities we have approximated $\theta_{\rm max} = \sin^{-1}\sqrt{R_{\rm ns}/R_{\rm Y}} \approx \sqrt{R_{\rm ns}/R_{\rm Y}}$ and have assumed $R_{\rm ns} = 12$ km.  

A numerical estimate of the total mass loss rate $\dot{M}_{\rm tot}$ integrated across the open zone (eq.~[\ref{eq:mdottot}]) for each neutrino luminosity and spin period are given in Table \ref{table:integrated}.  The value of $\dot{M}_{\rm tot}$ varies from a factor of $\sim 10$ times larger to $\sim 30$ times smaller than the spherical case as the spin period increasees from $P = 1$ ms to 10 ms.  This transition occurs as the centrifugal enhancement at small $P$ is offset by the shrinking open fraction $f_{\rm open} \propto P^{-1}$ for large $P$.  A larger value of $\dot{M}_{\rm tot}$, by a factor $\gtrsim R_{\rm L}/R_{\rm Y} \sim 3$, will result if the closed zone is larger $R_{\rm Y} \sim R_{\rm L}/3$ than that predicted under our assumption that $R_{\rm Y} = R_{\rm L}$ ($\S\ref{sec:forcefree}$; Fig.~\ref{fig:hydrostatic}).  
\subsubsection{Final entropy} 
\label{sec:entropy}

The asymptotic values of the entropy for our fiducial rotating solutions are $S = 87$ $k_{\rm b}$ n$^{-1}$ and $51$ $k_{\rm b}$ n$^{-1}$ for the $\theta = 0$ and $\theta = \theta_{\rm max}$, respectively, as compared to $S = 87$ $k_{\rm b}$ n$^{-1}$ for the equivalent spherical case.  These similarities and differences can be understood from the fact that the entropy increase of the outflowing matter is approximately given by
\be
 \De S \propto \int \frac{dQ}{T}\approx \frac{\De Q}{\bar{T}} \label{eq:DeltaS},
\ee
where $dQ$ is the net heating per nucleon and $\bar{T}$ is the average temperature in the heating region.  To first order, the entropy of the polar rotating outflow is similar to that in the spherical case for the same reason that $\dot{M}$ is similar: heating is concentrated in the hydrostatic region and the value of $\Delta Q$ is essentially fixed by that required to escape the gravitational well of the PNS.  In the case of inclined field lines $\theta = \theta_{\rm max}$ the entropy is reduced because less neutrino heating $\De Q$ is necessary to escape the potential well given the additional source of centrifugal acceleration.  The value of $S$ in this case is nevertheless still significantly larger than for an equatorial outflow in the monopolar case with the same neutrino luminosity and spin period \citep{Metzger+07}.  

Figure \ref{fig:svstautwoms} shows the asymptotic entropy across a range of PNS properties.  The strongest trend is that entropy  decreases with increasing neutrino luminosity, approximately as $S \propto L_{\nu}^{-1/6}\langle \epsilon_{\nu}\rangle^{-1/3} \propto L_{\nu}^{-1/4}$ (\citealt{Qian&Woosley96}).  This dependence can also be understood from equation (\ref{eq:DeltaS}): the mean temperature in the region of net heating is smaller for lower neutrino luminosities because $\bar{T}$ is to first order set by the balance between neutrino heating $\propto L_{\nu}$ and neutrino cooling $\propto T^{6}$.  

\subsubsection{Expansion timescale}
\label{sec:texp}
Matter flows across the range of radii at which alpha particles form on the expansion timescale \citep{Hoffman+97}
\be
 t_{\rm exp}=1.28\left.\frac{1}{v}\frac{d \ln T}{dl}\right|_{\rm T = 0.5 MeV}
\label{eq:texp}
\ee
The expansion timescale for our fiducial rotating solutions are $t_{\rm exp} = 11$ ms  and $0.6$ ms for the $\theta = 0$ and $\theta = \theta_{\rm max}$, respectively, as compared to $t_{\rm exp} = 13$ ms for the equivalent spherical case (Fig.~\ref{fig:multitherm}; {\it bottom panel}).  Faster expansion in the magnetized case results from more rapid acceleration (Fig.~\ref{fig:multidyn}; {\it top panel}) due to the faster divergence of the dipolar areal function (Fig.~\ref{fig:Br}) and the related fact that for similar entropy a faster decrease in density results in $T = 0.5$ MeV being achieved closer to the PNS surface, where the radius (and hence expansion time) is likewise smaller (Fig.~\ref{fig:multidyn}; {\it bottom panel}).  

The expansion time is further reduced by a large factor for $\theta > 0$ due to centrifugal acceleration, which increases the outflow velocity as compared to the purely thermally-driven case.  The value of $t_{\rm exp}$ in Table \ref{table:summary} shows that at large angles $\theta \approx \theta_{\rm max}$ centripetal acceleration is more important than the purely geometric effect of areal divergence for spin periods $P\le 4$ ms.  For our fiducial solution, for instance, areal divergence reduces $t_{\rm exp}$ from $13$ ms for spherical solution to $11$ ms for polar outflows ($\theta = 0$).  In contrast, centrifugal acceleration decreases the dynamical timescale from 11 ms for polar outflow to 0.6 ms for outflows along the last open field line.

Figure \ref{fig:svstautwoms} compares $t_{\rm exp}$ for solutions with different rotation rates and neutrino luminosities.  The strongest trends are that (1) $t_{\rm exp}$ decreases with increasing neutrino luminosity (as also occurs in spherical, non-rotating winds; \citealt{Thompson+01}); (2) at fixed neutrino luminosity, $t_{\rm exp}$ is smaller for shorter spin periods $P$ and larger outflow angles $\theta$, again due to centrifugal acceleration.     

\begin{figure}
\subfigure{
\includegraphics[width=0.5\textwidth]{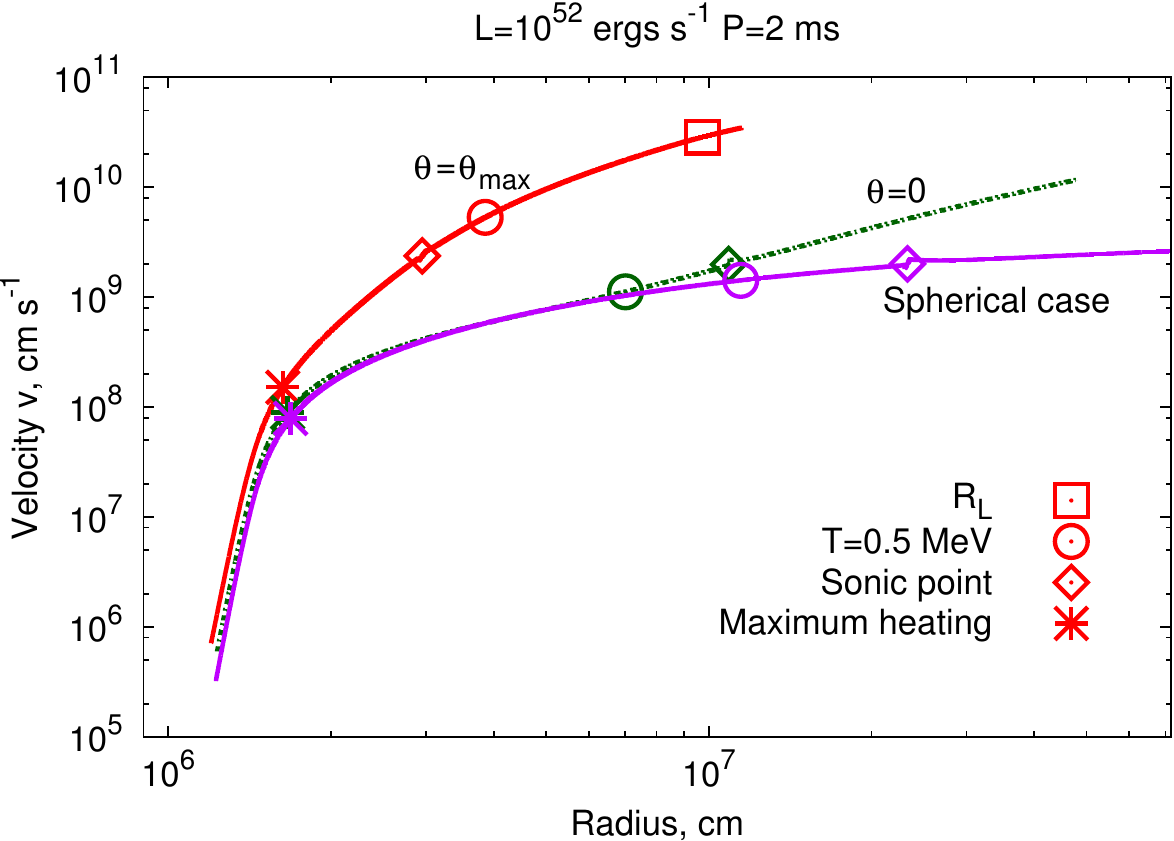}}
\subfigure{
\includegraphics[width=0.5\textwidth]{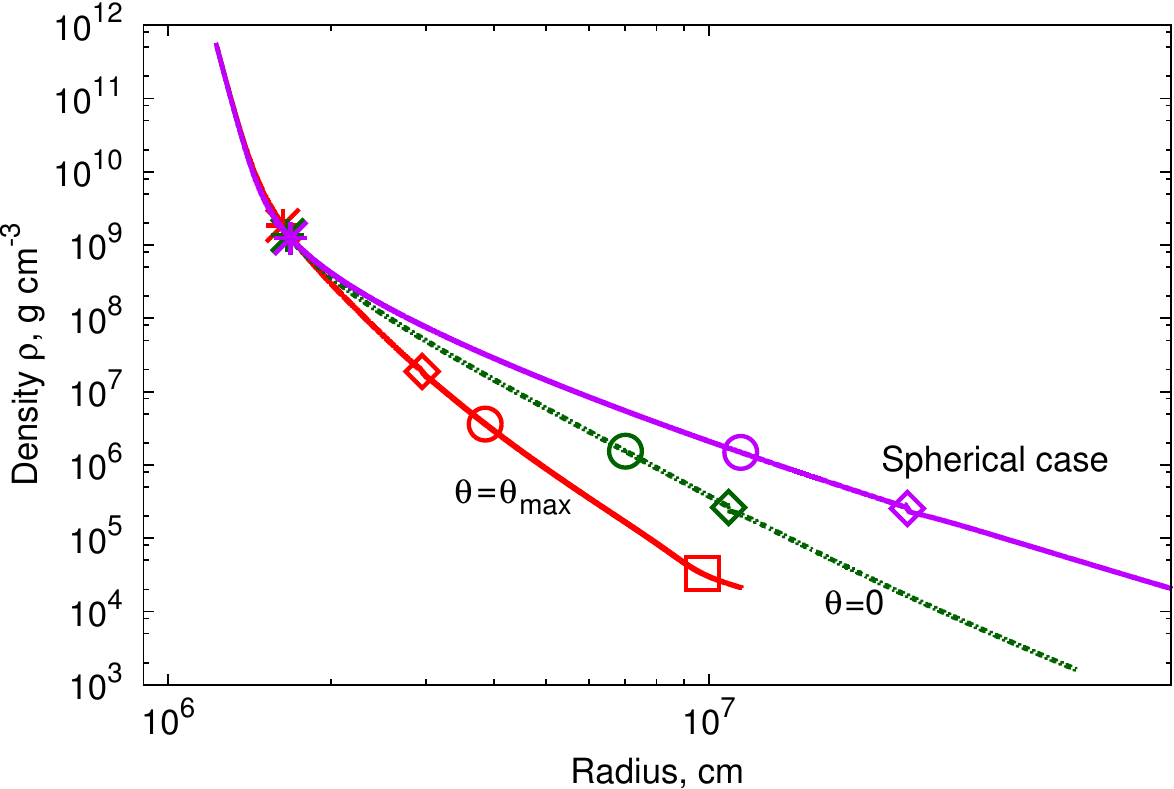}} 
\subfigure{
\includegraphics[width=0.5\textwidth]{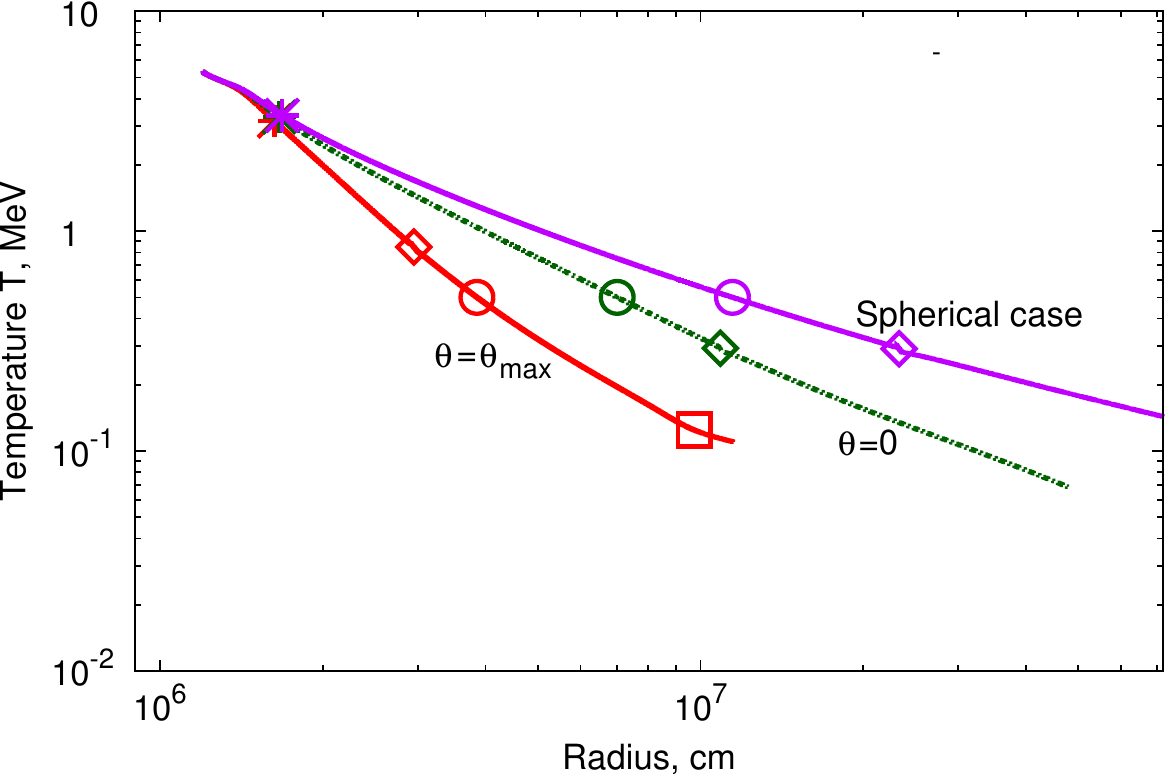}} 
\caption{Radial profiles of velocity $v$ ({\it top}), density $\rho$ ({\it middle}), and temperature $T$ ({\it bottom}) for our fiducial solution with $L_{\nu} = 10^{52}$ erg s$^{-1}$ and $P = 2$ ms, shown for different field lines corresponding to a polar outflow ($\theta = 0$), maximally inclined outflow ($\theta = \theta_{\rm max}$).  Shown for comparison is a spherical, non-rotating outflow for the same neutrino luminosity.  Critical radii are marked with symbols, including the light cylinder $R_{\rm L}$ ({\it square}), radius of $\alpha-$particle formation $R_{\rm 0.5Mev}$ ({\it circle}), sonic point ({\it diamond}), and point of maximum net heating ({\it asterisk}). } 
\label{fig:multidyn}
\end{figure}

\begin{figure}
\subfigure{
\includegraphics[width=0.5\textwidth]{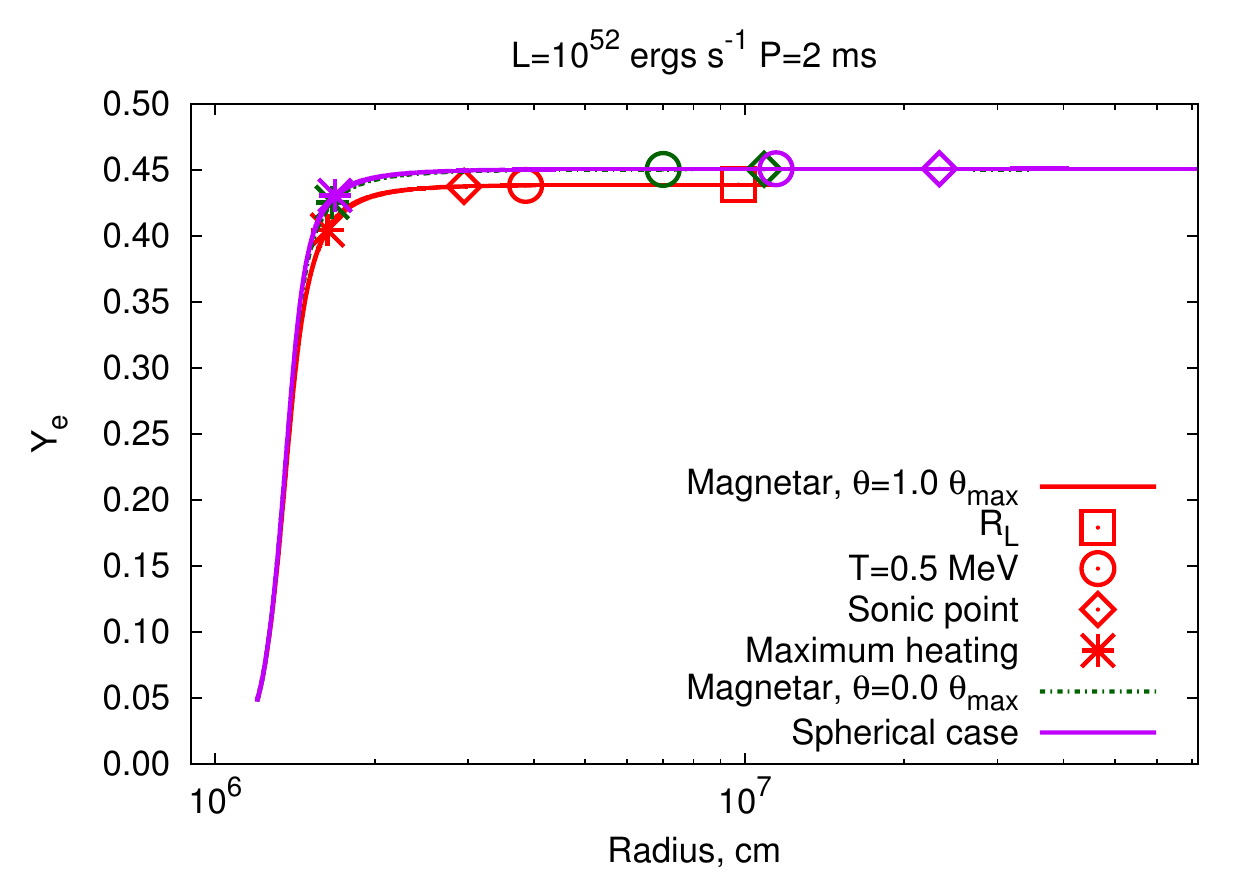}} 
\subfigure{
\includegraphics[width=0.5\textwidth]{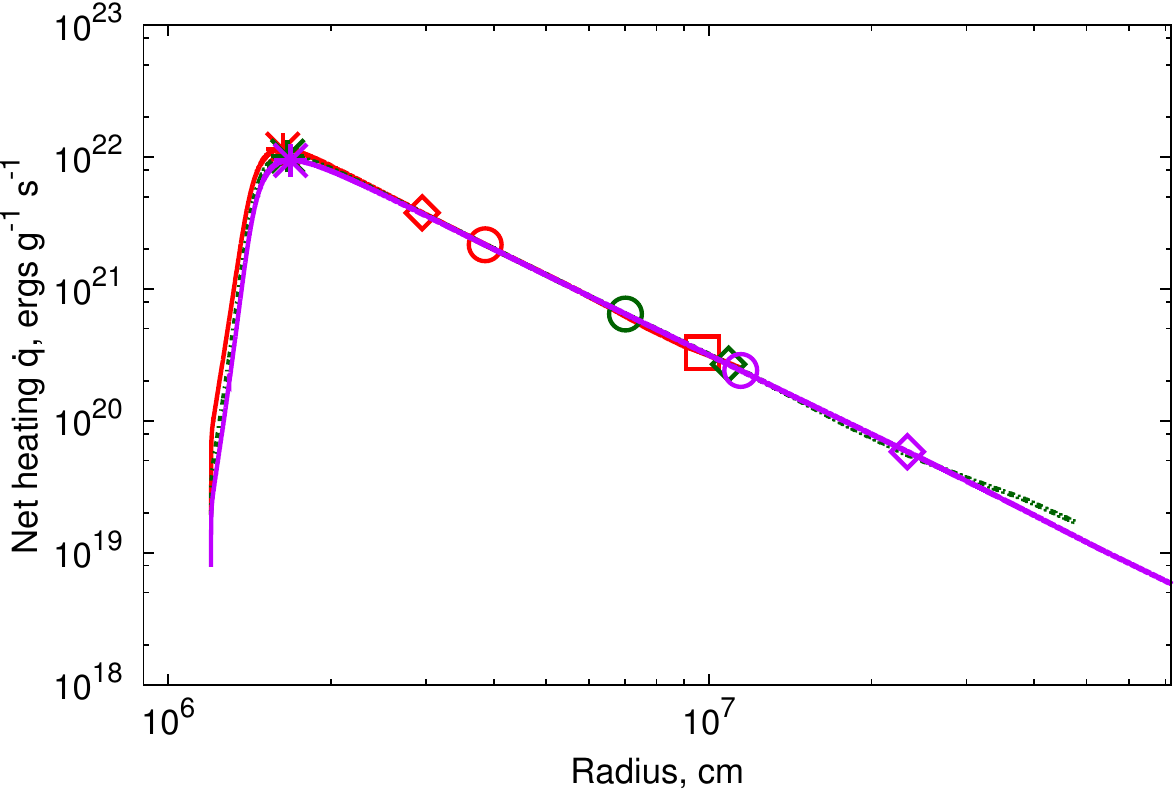}} 
\subfigure{
\includegraphics[width=0.5\textwidth]{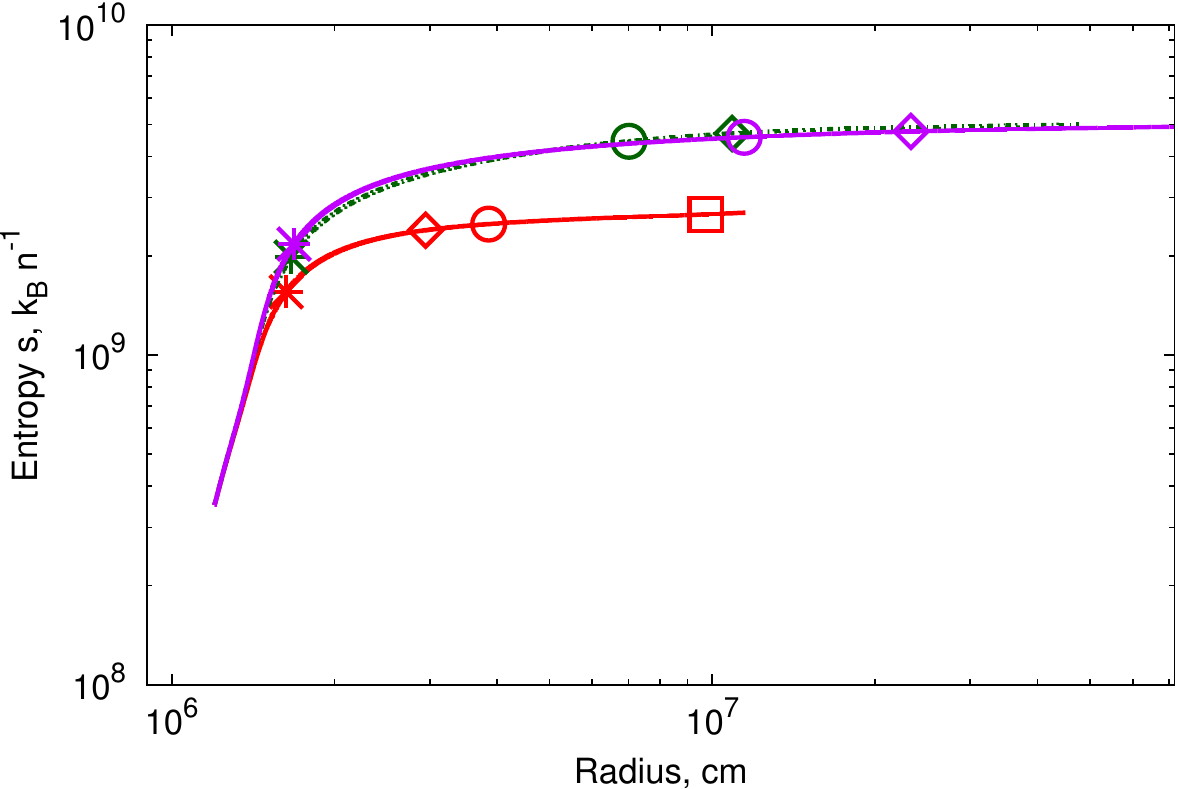}} 
\caption{Radial profiles of the electron fraction $Y_e$ ({\it top}), net heating rate $\dot{q}$ ({\it middle}), and entropy s ({\it bottom}) for our fiducial solution with $L_{\nu} = 10^{52}$ erg s$^{-1}$ and $P = 2$ ms (same as Fig.~\ref{fig:multidyn}).  Radii marked with symbols are the same as in Figure \ref{fig:multidyn}.
} 
\label{fig:multitherm}
\end{figure}

\begin{figure}
\begin{center}
\includegraphics[width=1.0\linewidth]{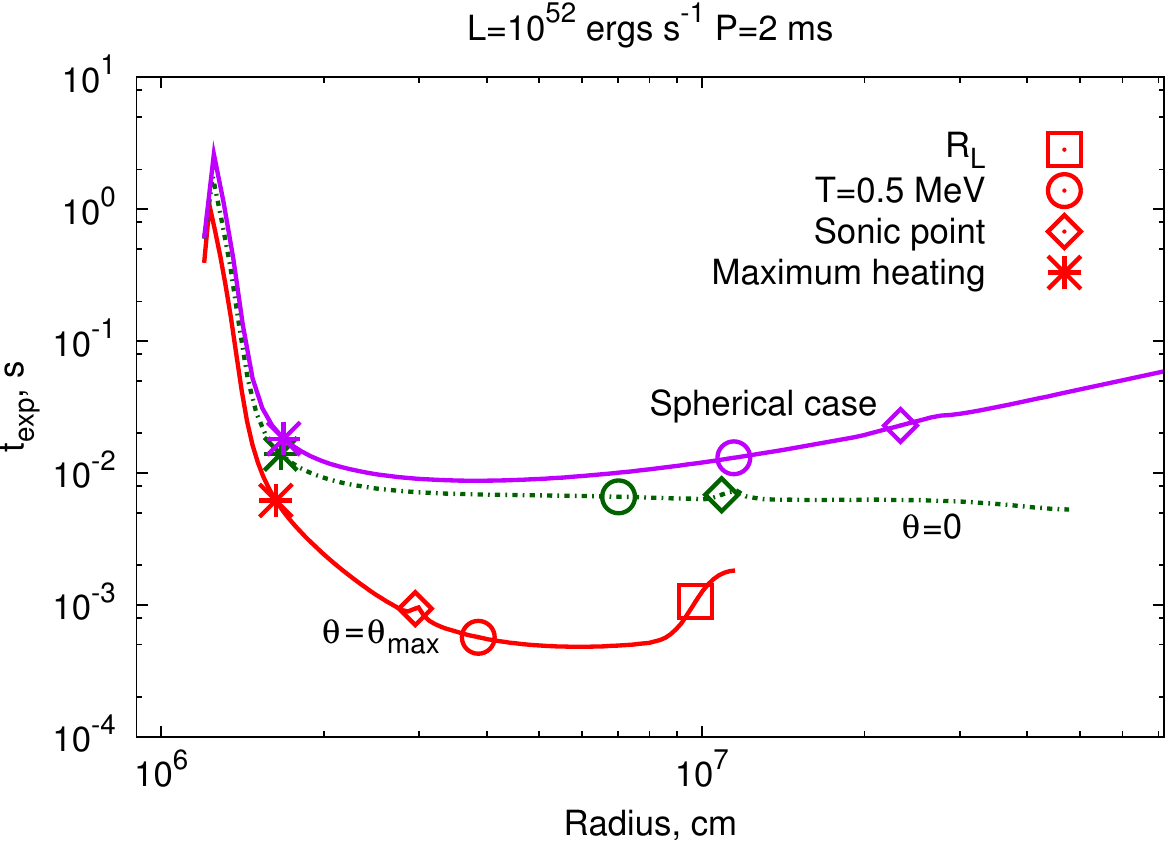}
\caption{Radial profiles of the local expansion timescale $t_{\rm exp}$ (eq.~[\ref{eq:texp}]) for our fiducial solution with $L_{\nu} = 10^{52}$ erg s$^{-1}$ and $P = 2$ ms, shown for different field lines corresponding to a polar outflow ($\theta = 0$), maximally inclined outflow ($\theta = \theta_{\rm max}$).  Shown for comparison is a spherical, non-rotating outflow for the same neutrino luminosity. }
\label{fig:tau}
\end{center}
\end{figure}

\begin{figure}
\begin{center}
\includegraphics[width=1.0\linewidth]{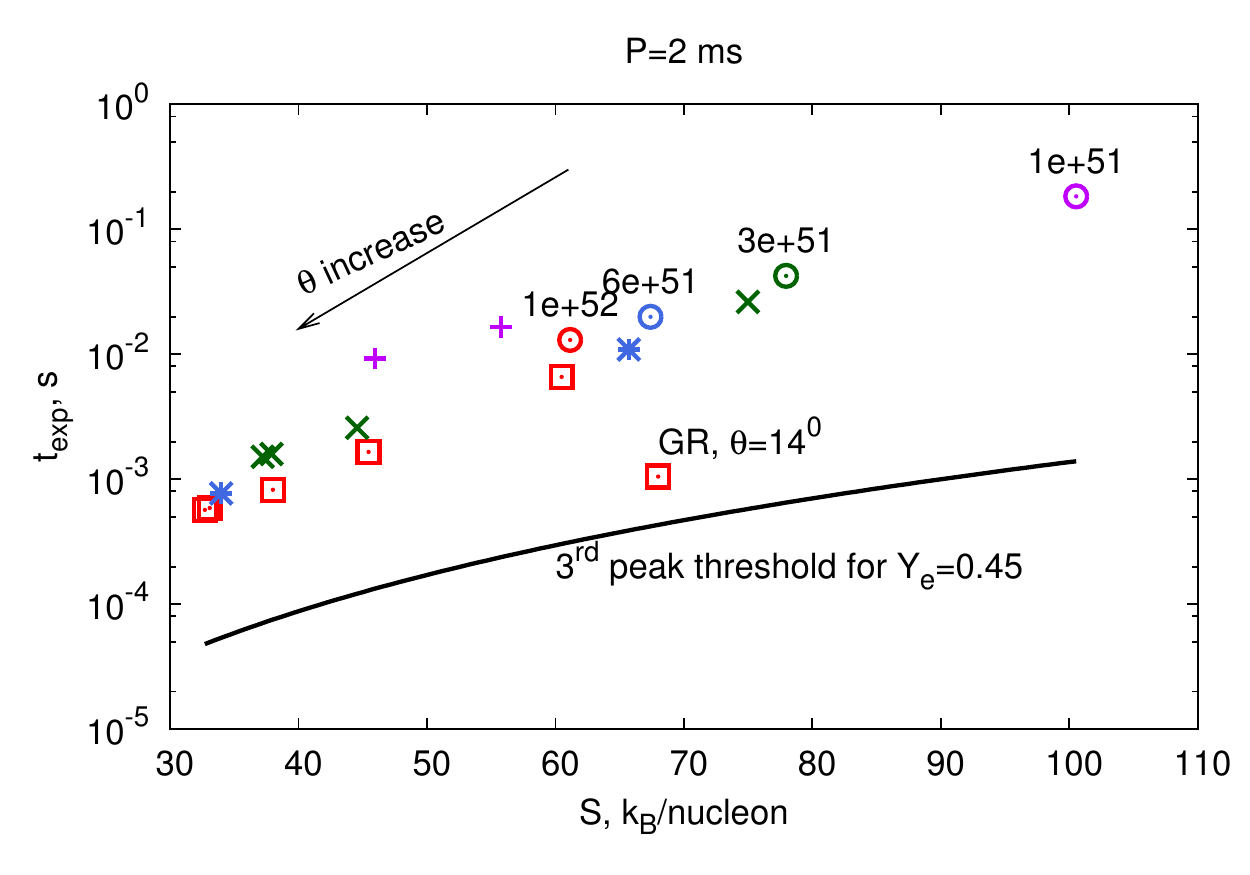}
\caption{Expansion timescale $t_{\rm exp}$ (eq.~[\ref{eq:texp}]) versus asymptotic entropy $S$, calculated for a PNS with spin period $P = 2$ ms and shown for different neutrino luminosities: $L_{\nu} = 10^{52}$ erg s$^{-1}$ ({\it red square}), $6\times 10^{51}$ erg s$^{-1}$ ({\it blue asterisk}), $3\times 10^{51}$ erg s$^{-1}$ ({\it green cross}), and $10^{51}$ erg s$^{-1}$ ({\it purple plus}).  Different field line angles $\theta$ are shown for each luminosity, with $\theta$ increasing to the lower left hand corner of the plot.  Circles of the same color show the corresponding equivalent spherical, non-rotating solutions with the same neutrino luminosity.  The solid black line shows the threshold for a successful third peak $r$-process $\eta = \eta_{\rm thr}$ (for an assumed electron fraction $Y_e = 0.45$).}
\label{fig:svstautwoms}
\end{center}
\end{figure}

\begin{figure}
\begin{center}
\includegraphics[width=1.0\linewidth]{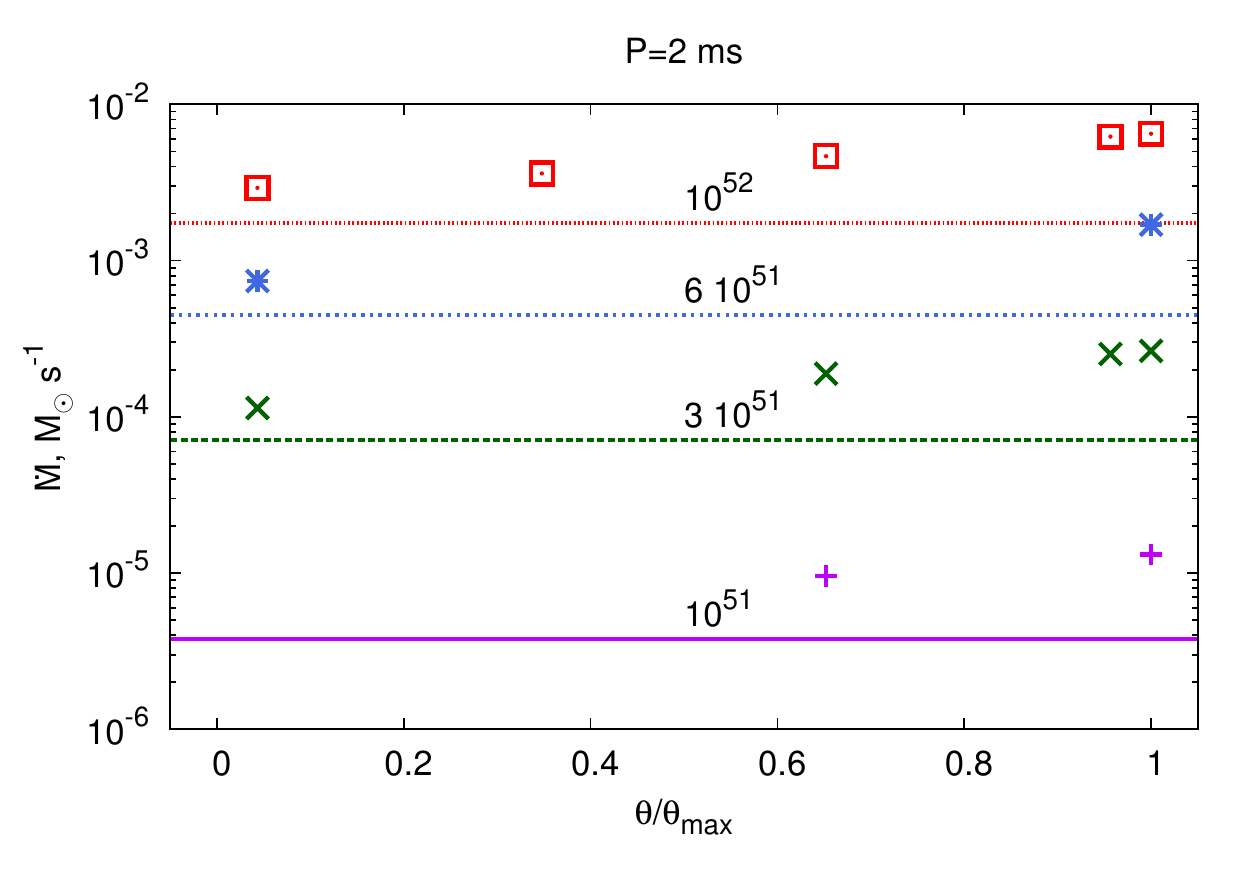}
\caption{Spherically equivalent mass loss $\dot{M}=4\pi r_0^2\rho_0v_0$ (eq.~[\ref{eq:Mdot}]) as a function of the field polar angle, calculated for a PNS with spin period $P=2$ ms for different neutrino luminosities.  Shown for comparison with horizontal lines are the mass loss rates of the spherical solutions of the same luminosities. Luminosities in ergs s$^{-1}$ are written near each line.
 }
\label{fig:mdottwoms}
\end{center}
\end{figure}

\begin{figure}
\begin{center}
\includegraphics[width=1.0\linewidth]{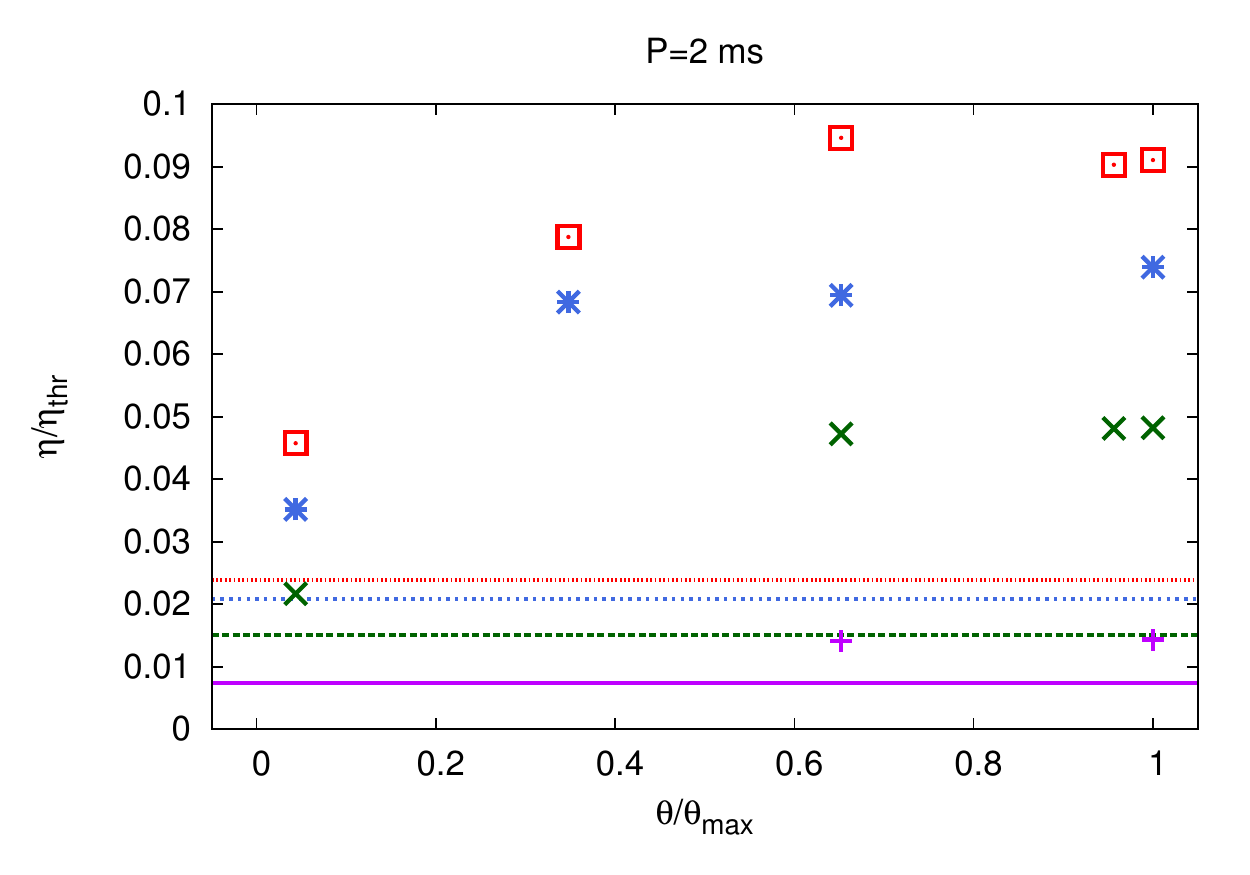}
\caption{Critical ratio $\eta \equiv S^{3}/Y_{e}^{3}t_{\rm exp}$ of wind properties for a successful $r$-process in units of the required threshold value $\eta_{\rm thr}$ (eq.~[\ref{eq:eta}]) as a function of the field polar angle, calculated for a PNS with spin period $P = 2$ ms for different neutrino luminosities: (from top to bottom) $L_{\nu} = 10^{52}$ erg s$^{-1}$ ({\it red square}), $6\times 10^{51}$ erg s$^{-1}$ ({\it blue asterisk}), $3\times 10^{51}$ erg s$^{-1}$ ({\it green cross}), and $10^{51}$ erg s$^{-1}$ ({\it green cross}).  Note that GR will act to enhance $\eta/\eta_{\rm thr}$ by a factor $\sim 3-4$ over the values shown. Shown for comparison with horizontal lines are the values of $\eta$ of the spherical solutions of the same luminosities. 
}
\label{fig:sctautwoms}
\end{center}
\end{figure}

\section{Discussion}
\label{sec:discussion}

\subsection{$r$-Process Nucleosynthesis}
\label{sec:rprocess}
Free nuclei recombine into $\alpha$-particles once the temperature decreases to $T \sim 5\times 10^{9}$ K, as our solutions show occurs several tens of kilometers above the PNS surface.  Heavier elements start to form once the temperature decreases further, via the reaction $^{4}$He($\alpha$n,$\gamma$)$^{9}$Be($\alpha$,n)$^{12}$C.  After $^{12}$C forms, additional $\alpha$ captures produce heavy `seed' nuclei with characteristic mass $\bar{A} \simeq 90-120$ and charge $\bar{Z}$ (\citealt{Woosley&Hoffman92}).  The $r$-process occurs as remaining free neutrons are captured onto these seed nuclei.  The maximum mass $A_{\rm max}$ to which the $r$-process proceeds depends on the ratio of free neutrons to seed nuclei following completion of the $\alpha$-process.  Because $^{12}$C production is the rate-limiting step to forming seeds, the neutron to seed ratio in turn depends on the electron fraction $Y_e$, entropy $S$, and expansion timescale $t_{\rm exp}$ (eq.~[$\ref{eq:texp}$]) of the outflow (\citealt{Meyer&Brown97}).  

For $Y_e \gtrsim \bar{A}/\bar{Z} \approx 0.35-040$ the condition for $r$-process to reach the third mass peak ($A_{\rm max} \gtrsim 190$) can be expressed as (\citealt{Hoffman+97}):
\be
\eta \equiv \frac{S^{3}}{Y_e^{3}t_{\rm exp}} \gtrsim \eta_{\rm thr} \approx 8\cdot 10^9\, \left(k_B {\rm n}^{-1}\right)^3 s^{-1},
\label{eq:eta}
\ee
where the ratio $\eta/\eta_{\rm thr}$ thus serves as a `figure of merit' for the potential success of a given $r$-process site.  Previous studies of the $r$-process in spherically symmetric, non-rotating PNS winds typically find that $\eta \ll \eta_{\rm thr}$ (e.g.~\citealt{Thompson+01}), thus disfavoring such events as sources of heavy $r$-process nuclei. 

Figure \ref{fig:sctautwoms} shows $\eta/\eta_{\rm thr}$ calculated for our $P = 2$ ms solutions corresponding to different field lines and neutrino luminosities.  Figure \ref{fig:sctauint} shows the (mass flux weighted) average value of $\eta/\eta_{\rm thr}$ over the entire outflow for different periods and neutrino luminosities.  Shown for comparison in each case are the values of $\eta/\eta_{\rm thr}$ for our spherical solutions of the same luminosities. 

\begin{figure*}
\begin{center}
\includegraphics[width=1.0\linewidth]{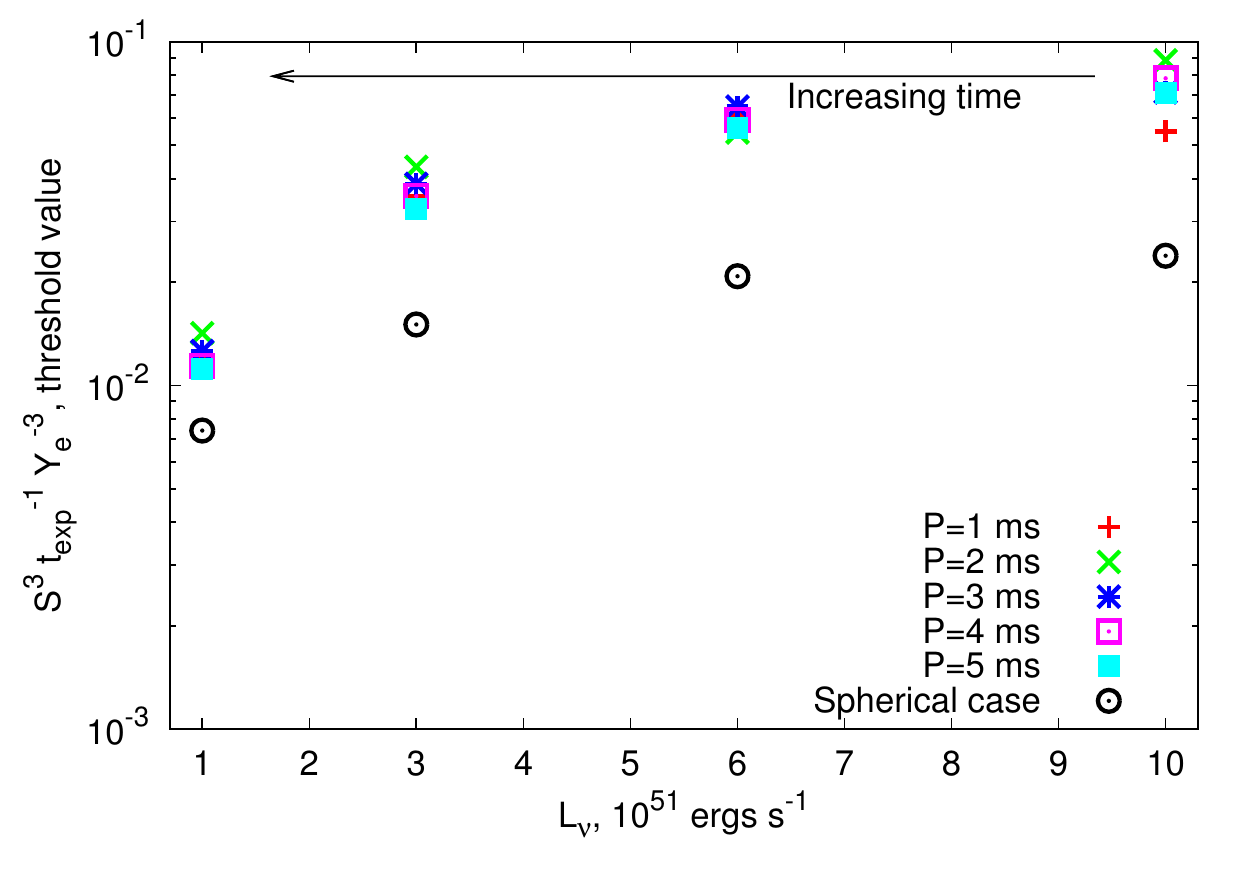}
\caption{Critical ratio $\eta \equiv S^{3}/Y_{e}^{3}t_{\rm exp}$ of wind properties required to achieve third-peak $r$-process in units of the threshold value $\eta_{\rm thr}$ (eq.~[\ref{eq:eta}]), mass-averaged over the open field lines as a function of the PNS neutrino luminosity for different rotation periods  $P = 1$ ms ({\it plus}), 2 ms ({\it cross}), 3 ms ({\it asterisk}), 4 ms ({\it open square}), and 5 ms ({\it filled square}).  Shown for comparison with dot-centered circles are the equivalent spherical wind solution for the same neutrino luminosity.  Note that GR will act to enhance $\eta$ by a factor $\sim 3-4$ over the values shown.}
\label{fig:sctauint}
\end{center}
\end{figure*}  
  
 Our most promising solutions for a PNS of mass 1.4 $M_{\odot}$ are for $L_{\nu} =10^{52}\,{\rm ergs\,s^{-1}}, \theta \approx \theta_{\rm max}$ and spin periods in the range $P \sim 2-5$ ms, for which $\eta$ is $\approx 10$ per cent of the threshold value $\eta_{\rm thr}$.  Although these solutions are still well below the threshold, they are a factor $\gtrsim 4$ times higher than the spherical, non-rotating wind solution of the same neutrino luminosity.  This enhancement results from the combination of a much lower value of $t_{\rm exp}$ due to centrifugal flinging and the dipolar geometry ($\S\ref{sec:texp}$) and an entropy that is only moderately smaller (by a factor $\sim 50\%$) than the spherical case. High entropy is maintained because the vertically directed geometry of the magnetic field near the surface provides matter with sufficient time in the heating region prior to onset of significant centrifugal acceleration at larger cylindrical radii ($\S\ref{sec:entropy}$).  Consistent with the results of previous studies (e.g.~\citealt{Otsuki+00}; \citealt{Thompson+01}; \citealt{Wanajo13}), we find that a larger PNS mass $M = 2M_{\odot}$ increases the value of $\eta$ by an additional factor of $\sim 4$ (Tables \ref{table:2Msun}, \ref{table:integrated2Msun}) as compared to the $M = 1.4M_{\odot}$ case, corresponding to approximately 40 per cent of the required threshold in our most promising case.  

In light of our conclusion that magnetic fields and rotation enhance the prospects for the $r$-process in PNS winds, we note several caveats.  First, our calculations do not include general relativity (GR).  Past studies of neutrino-heated winds have found that GR effects increase the final entropy\footnote{Higher entropy results primarily from the deeper effective gravitational potential well of the PNS in GR, which requires a larger amount of neutrino heating $\De Q$ to escape, thus increasing $S$ according to equation (\ref{eq:DeltaS}).} and decrease the dynamical timescale.  If we adopt the $\sim 50\%$ increase in entropy found by \citet{Thompson+01} (see also \citealt{Cardall&Fuller97}; \citealt{Otsuki+00}), then the resulting increase in $\eta$ by a factor of 3 is nearly sufficient to push our most promising models for $M = 1.4 M_{\odot}$ to success (and more than sufficient for $M = 2M_{\odot}$), and to place our less promising models to within a factor of $\sim 2-3$ of the threshold.  In order to explore the effects of GR further, our most promising solution $L_{\nu} =10^{52}\,{\rm ergs\,s^{-1}}, P=2\,{\rm ms}, \theta=0.7\theta_{\rm max}$ was recalculated with a \citet{Paczynsky&Wiita80} gravitational potential.  The value of $\eta$ of the corresponding solution was found to be $\sim 4$ times higher than in the Newtonian case (see Tables \ref{table:summary}, \ref{table:2Msun}), which for $M = 2M_{\odot}$ is above the third peak $r$-process threshold.  

Another caveat is that we assume a small PNS radius of $12$ km, even at the highest neutrino luminosities corresponding to the earliest times ($t \sim 1$ s) in the PNS cooling evolution.  At such early stages after the explosion, the PNS is inflated by thermal pressure support relative to its final size (e.g.~\citealt{Pons+99}).  A larger PNS radius results in a lower value of $\eta$ than would be calculated assuming a fully contracted PNS (the rotation rate prior to full contraction is also likely to be lower due to angular momentum conservation) due to the shallower gravitational potential well (eq.~[\ref{eq:DeltaS}]).  We note, however, that \citet{Pons+99} do not include the effects of convective cooling, which may act to enhance the rate of PNS cooling and contraction (\citealt{Scheck+06}; \citealt{Roberts12}).  

A final caveat is that, although $\eta$ is a useful proxy for the success of the $r$-process, a full reaction network calculation is necessary to more precisely quantify the nucleosynthetic yield of magnetar birth.  We briefly comment on aspects of the nucleosynthesis that may differ from standard spherical, non-rotating PNS winds.  Due to the rapid expansion of the outflow, $\alpha$-particle formation occurs much closer to the PNS surface than in the spherical case.  Although in principle this could imply a greater role for neutrino reactions on the process of weak freeze-out itself, we find that the faster outflow expansion negates this effect.  In particular the product of the neutrino flux times the expansion timescale (neutrino fluence) at the $\alpha$ formation radius is found for all solutions to be similar to that for the non-rotating, spherical wind with the same neutrino luminosity.  



\subsubsection{Proto-magnetars as Galactic $r$-process site}

We shall proceed under the assumption that the $r$-process is indeed successful in proto-magnetar winds with moderately rapid rotation $P \sim$ few ms, due to the inclusion of GR and other potential effects not taken into account, and consider the resulting implications.  Our most promising models have luminosities $L=10^{52}$ ergs s$^{-1}$  and $P \lesssim 5$ ms.

Producing all of the heavy $r$-process elements in the Galaxy over its lifetime requires a mean production rate $\dot{M}_{\rm r} \sim 5\times 10^{-7}M_{\odot}$ yr$^{-1}$ (\citealt{Qian00}), thus requiring an ejecta mass per event of
\be
\bar{M}_{\rm ej} = \frac{\dot{M}_{\rm r}}{ f_{\rm r}\mathcal{R}} \approx 1.5\times 10^{-3}M_{\odot}\left(\frac{f_{\rm r}}{0.1}\right)^{-1}\left(\frac{\mathcal{R}}{0.1\mathcal{R}_{\rm CC}}\right)^{-1}
\label{eq:Mej}
\ee
for an event rate $\mathcal{R}$, which we have normalized to the estimated rate of core collapse supernovae of $\sim 3$ per century (\citealt{Adams+13}), where $f_{\rm r} \approx 1-2Y_e \sim 0.1-0.2$ is the fraction of the ejecta mass placed in $r$-process nuclei (as opposed to $^{4}$He).    

\begin{figure}
\begin{center}
\includegraphics[width=1.0\linewidth]{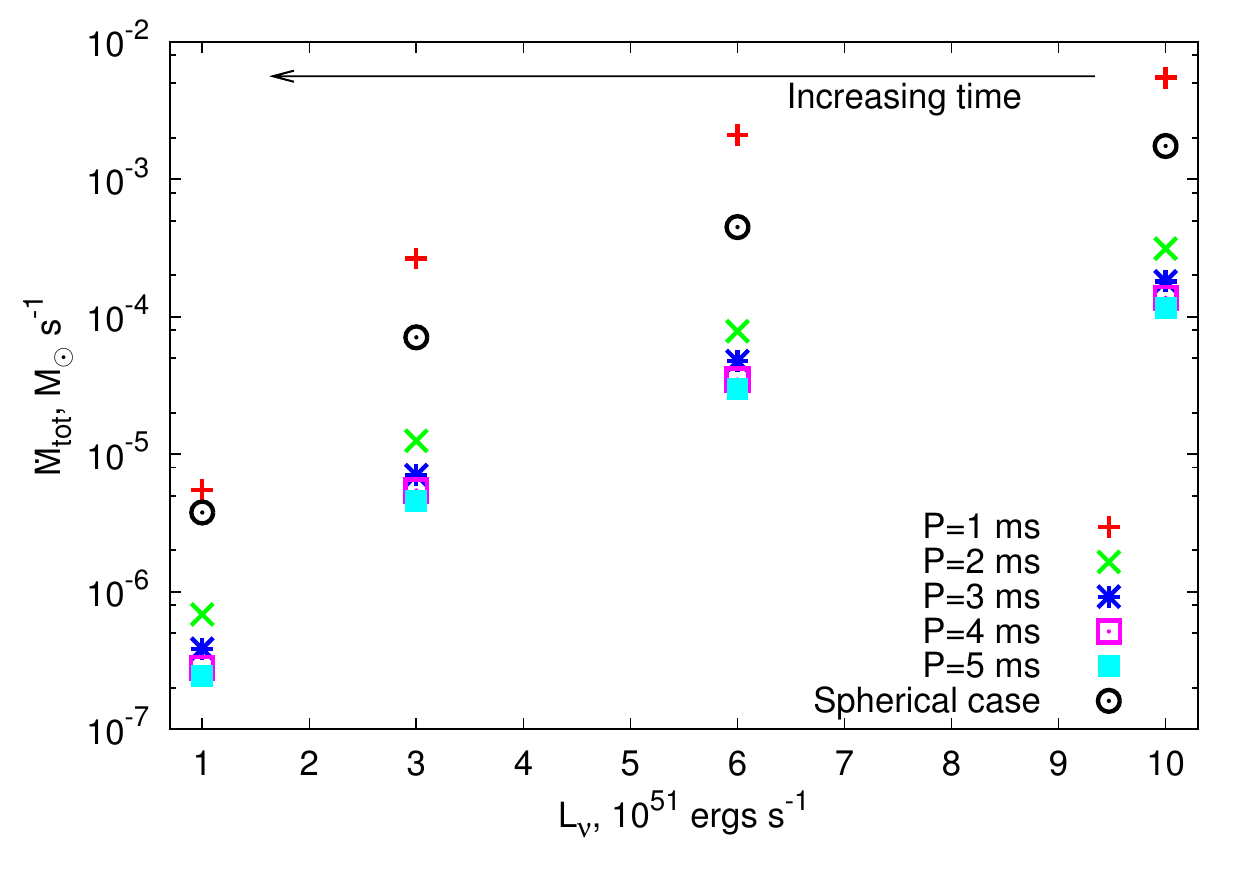}
\caption{Total mass loss rate $\dot{M}_{\rm tot}$ integrated over open field lines (eq.~[\ref{eq:mdottot}]) as a function of the PNS neutrino luminosity for different rotation periods $P = 1$ ms ({\it plus}), 2 ms ({\it cross}), 3 ms ({\it asterisk}), 4 ms ({\it open square}), and 5 ms ({\it filled square}).    Shown for comparison with dot-centered circles are the equivalent spherical wind solution for the same neutrino luminosity. }
\label{fig:mdotint}
\end{center}
\end{figure}

Our models with $P = 2-5$ ms have total ejecta masses $M_{\rm ej} \sim 10^{-4} M_{\odot}$ (Table \ref{table:integrated}), which could be enhanced by a factor $\gtrsim f_{\rm open} \propto  R_{\rm L}/R_{\rm Y} \sim 3$ (eq.~[\ref{eq:fopen}]) due to the finite pressure of the closed zone (eq.~[\ref{fig:hydrostatic}]) or due to coupling between the proto-magnetar outflow and the surrounding envelope of the star \citep{Bucciantini+07}.  A somewhat larger open zone (larger $\theta_{\rm max}$) could also act to enhance the $r$-process figure of merit $\eta$, because for $L_{\nu} = 10^{52}$ erg s$^{-1}$, $\eta$ is still increasing monotonically with $\theta$ for $P \gtrsim 3$ ms (Table \ref{table:summary}).

Combining these estimates for the ejecta with equation (\ref{eq:Mej}), we conclude that $\gtrsim 30\%$ of the Galactic $r$-process could in principle originate from proto-magnetars with birth periods $P < 5$ ms at a rate that is $\gtrsim 10\%$ of the core collapse rate (\citealt{Woods&Thompson06}).  Their contribution could be greater if the $r$-process receives contributions from extremely rapidly spinning neutron stars with $P = 1$ ms (which contribute higher $M_{\rm ej}$ but are probably much rarer; see also \citealt{Winteler+12}).  

A population of moderately rapidly-rotating magnetar $r$-process sources is compatible with a variety of other observations.  A rotation period $P \gtrsim 4$ ms corresponds to a total rotational energy $E_{\rm rot} \lesssim 2\times 10^{51}$ ergs, which is too low to be excluded as accompanying the birth of Galactic magnetars based on the energetics of their SN remnants (\citealt{Vink&Kuiper06}).  Magnetar birth, though likely not as common as normal core collapse supernovae (however, see \citealt{Rea+13}), is more common than binary neutron star mergers, the other commonly discussed $r$-process site.  An $r$-process source from magnetar birth would thus be in better agreement with studies of Galactic chemical evolution that appear to disfavor infrequent, high-$M_{\rm ej}$ events (e.g.~\citealt{Argast+04}; however, see \citealt{Tsujimoto&Shigeyama14}).   

Magnetar birth in core collapse supernovae could also provide prompt enrichment, as would accompany even the earliest stages of Galactic chemical evolution, another observation in tension with the NS merger scenario.  The discovery of magnetars in massive clusters suggest that they result from the core collapse of very massive stars $\gtrsim 40M_{\odot}$ (\citealt{Figer+05}; \citealt{Muno+06}; \citealt{Bibby+08}), which are expected to produce massive neutron stars more favorable to the $r$-process (Table \ref{table:2Msun}, \ref{table:integrated2Msun}; \citealt{Wanajo13}).  On the other hand, massive progenitors may be disfavored by the lack of correlation between Fe production and third peak $r$-process elements based on the observed abundances in metal-poor stars (e.g.~\citealt{Qian&Wasserburg03}).  Finally, the birth of a millisecond magnetar is one of the leading theoretical models for explaining recently discovered class of hydrogen-poor `super-luminous' supernovae (\citealt{Kasen&Bildsten10}; \citealt{Woosley10}; \citealt{Metzger+14}), events which are indeed observed to occur in very metal poor galaxies (e.g.~\citealt{Lunnan+14}) similar to those encountered in the early history of the Galaxy.

\subsection{Gamma-Ray Burst Outflows}
\label{sec:GRB}
\begin{figure}
\begin{center}
\includegraphics[width=1.0\linewidth]{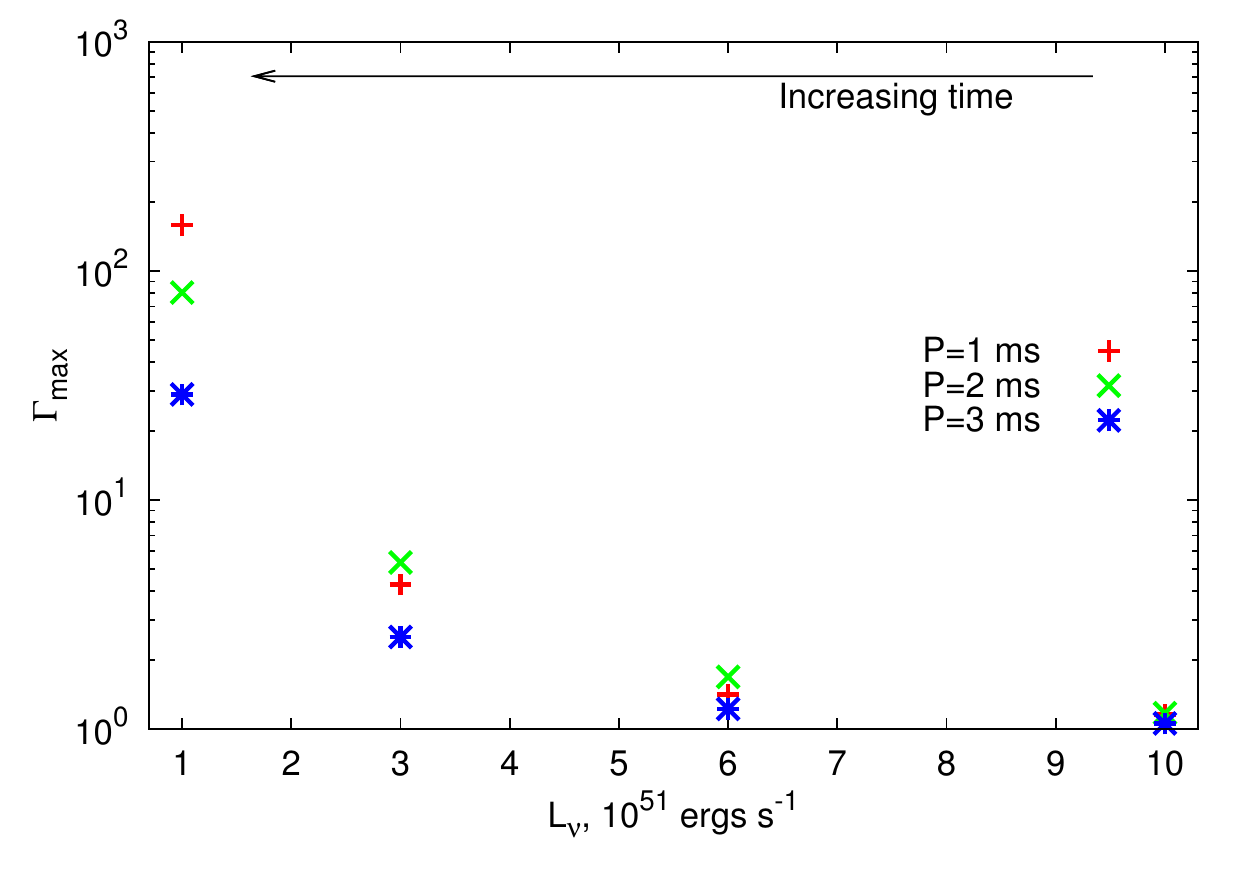}
\caption{Maximum Lorentz $\Gamma_{max}$ achieved by proto-magnetar outflows as a function of neutrino luminosity calculated according to equation (\ref{eq:gamma}]) for an assumed surface field strength $B =  3\times 10^{15}$ G and based on the baryon loading determined from our wind solutions, shown for spin periods $P = 1$ ms ({\it plus}), 2 ms ({\it cross}), and 3 ms ({\it asterisk}).  Lower neutrino luminosities correspond to later times following core bounce, indicating that proto-magnetars with $P \lesssim 2$ ms achieve $\Gamma_{\rm max} \sim 100-1000$ on timescales of several seconds.
}
\label{fig:gamma}
\end{center}
\end{figure}
\subsubsection{Lorentz Factors (Baryon Loading)}
Millisecond proto-magnetars are considered promising central engines for powering gamma-ray bursts (GRBs) (e.g.~\citealt{Usov92}; \citealt{Wheeler+00}; \citealt{Thompson+04}; \citealt{Metzger+11a}).  One striking feature of GRB outflows are their large Lorentz factors $\sim 100-1000$ (\citealt{Lithwick&Sari01}), which require a specific amount of entrained baryonic mass (`baryon loading').  

In the proto-magnetar model the GRB jet is powered by the electromagnetic extraction of the rotational energy of the PNS, as occurs at the rate
\be
\dot{E} = \frac{\mu^{2}\Omega^{4}}{c^{3}} \approx 4\times 10^{49}\left(\frac{B_0}{10^{15}{\rm G}}\right)^{2}\left(\frac{P}{\rm ms}\right)^{-4}\,{\rm erg\,s^{-1}}
\label{eq:edot}
\ee
where $\mu = B_{0}R_{\rm ns}^{3}/2$ is the magnetic dipole moment.  The maximum Lorentz factor that the outflow can achieve at a given time is set by the ratio of the outflow power to its rest mass flux
\be
\sigma = \Gamma_{\rm max} = \frac{\dot{E}+\dot{M}_{\rm tot}c^{2}}{\dot{M}_{\rm tot}c^{2}}.
\label{eq:gamma}
\ee
We have assumed $R_{\rm Y} = R_{\rm L}$ in equation (\ref{eq:edot}), but $\Gamma_{\rm max}$ will be moderately higher if the open zone is larger ($R_{\rm Y} \lesssim R_{\rm L}$) because to first order $\dot{M}$ is proportional to the open magnetic flux (neglecting centrifugal enhancements), while $\dot{E}$ is proportional to the open flux {\it squared}.  

Figure \ref{fig:gamma} shows $\Gamma_{\rm max}$ calculated using $\dot{M}_{\rm tot}$ from our solutions with $P = 1-3$ ms (Fig.~\ref{fig:mdotint}; Table \ref{table:integrated}) as a function of neutrino luminosity for an assumed magnetic field $B_0 = 3\times 10^{15}$ G (this is the field strength required to produce a jet with power according to eq.~[\ref{eq:edot}] similar to those of observed GRBs).  Our results show that at high luminosities, corresponding to early times after core bounce, $\Gamma_{\rm max}$ is relatively small $\lesssim 10$.  However, at lower luminosities corresponding to later times of several seconds after core bounace, $\Gamma_{\rm max}$ increases to $\sim 100$.  Our results thus confirm previous findings (\citealt{Thompson+04}; \citealt{Metzger+07}; \citealt{Metzger+11a}) that proto-magnetars with short spin periods $P \sim 1-2$ ms are in principle capable of producing outflows with the correct range of Lorentz factors over the appropriate timescale to power long GRBs.  By contrast, slower rotating magnetars with $P \sim 3$ ms produce outflows with lower luminosities and $\Gamma_{\rm max} \sim$ 10, which could be responsible for less energetic or less relativistic phenomena such as low luminosity GRBs or X-ray flashes \citep{Metzger+11a}.     

\subsubsection{Ultra-High Energy Cosmic Rays}

A surprising discovery by the Pierre Auger Observatory (PAO) is that the composition of ultra high energy cosmic rays (UHECRs) appears to be dominated by heavy nuclei (with mass similar to iron) at the highest cosmic ray energies $\gtrsim 5\times 10^{19}$ eV (e.g.~\citealt{Abraham+10}; \citealt{Cazon+12}; however, see e.g.~\citealt{Sokolsky&Thomson07}).  Such a heavy-dominated composition is unexpected for most astrophysical sources such as AGN, but it could arise naturally if UHECRs are accelerated in GRB jets as a result of nucleosynthesis in proto-magnetar outflows (\citealt*{Metzger+11b}; \citealt{Horiuchi+12}).  

Previous sections have shown that proto-magnetar winds possess the necessary properties to synthesize heavy elements.  However, an important question is whether the composition of the jet will be dominated by helium and seed nuclei ($A \sim 90$), or whether the nucleosynthesis will have time to proceed to even heavier $r$-process nuclei.  If `ultra-heavy' nuclei contribute significantly to UHECRs, this would impact the cosmic ray energy spectrum, e.g. producing changes in the shape of the GZK cut-off (\citealt{Metzger+11b}).\footnote{The standard Greisen-Zatsepin-Kuzmin effect results from the energy loss of UHECRs due to pions produced when protons interact with the cosmic microwave background.  A similar energy loss process applies to heavy nuclei due to the loss of nuclei following excitation of giant dipole resonances by CMB photons (e.g.~\citealt{Hooper+05}; \citealt{Allard&Protheroe09}).}  Likewise, the abundance of free neutrons remaining in the jet has implications for the GRB emission process \citep{Beloborodov10} and the resulting neutrino emission (\citealt{Meszaros&Rees00}; \citealt{Gao&Meszaros12}; \citealt{Murase+13}; \citealt{Bartos+13}).

To check whether proto-magnetar powered GRB jets can in principle contain ultra-heavy nuclei, we must confirm that neutrons have sufficient time to capture onto seed nuclei in the rapidly expanding outflow.  The rate limiting step in capturing neutrons during the $r$-process are $\beta-$decays, which have typical half-lives $t_{\rm r} \sim$ seconds for the nuclei of interest (\citealt{Metzger+10}).  This timescale must be compared to that required for neutrons to be captured onto seed nuclei
\begin{eqnarray}
t_{\rm capt} &=& \frac{1}{n_{\rm n}\sigma v_{\rm th}} \nonumber \\
&\approx& 0.02{\rm \,s}\left(\frac{L_{\rm j}}{10^{50}\rm erg\,s^{-1}}\right)^{-1}\left(\frac{X_n}{0.1}\right)^{-1}\left(\frac{\sigma}{10^{-25}{\rm cm^{2}}}\right)^{-1}\left(\frac{T}{10^{7}\rm K}\right)^{-1/2}
\label{eq:tcapt}
\end{eqnarray}
where $n_{\rm n} = X_{\rm n}\rho/m_p$ is the neutron number density, $X_{\rm n}$ is the neutron mass fraction, $\rho = L_{\rm sd}/4\pi R_{\rm r}^{2}c^{3}$ is the jet density (assuming it remains spherical and mildly relativistic at radii $r \lesssim R_{\rm r}$), $R_{\rm r} \sim c t_{\rm r} \sim 10^{11}$ cm is the outflow radius on the timescale over which neutrons are captured, and $\sigma$ is the neutron capture cross section normalized to a characteristic value, and $v_{\rm th} \approx (kT/m_p)^{1/2}$ is the neutron thermal velocity.  Equation (\ref{eq:tcapt}) shows that for typical parameters the neutron capture timescale is shorter than the expansion timescale of the jet $\sim t_{\rm r}$, indicating that heavy $r$-process nuclei may indeed be expected in GRB jets given a sufficiently high abundance of neutrons relative to seed nuclei.  A more detailed calculation is necessary to determine whether a small fraction of free neutrons will avoid capture and hence remain in the jet to large radii.

\section{Conclusions}
\label{sec:conclusions}
We have calculated steady-state neutrino-heated outflows from magnetized, rotating proto-neutron stars (`proto-magnetars') under the assumption of an axisymmetric force-free dipolar geometry (Figs.~\ref{fig:schematic}, \ref{fig:Br}).  Our conclusions are summarized as follows:
\begin{itemize}
\item{The force-free approximation is justified in magnetized proto-neutron stars outflows at temperatures $T \gtrsim 0.5$ MeV relevant to heavy element nucleosynthesis for surface magnetic fields $B_0 \gtrsim 10^{14}-10^{15}$ G similar to those of Galactic magnetars (Fig.~\ref{fig:Bmin}) at neutrino luminosities achieved over the first $\sim 10$ seconds following core bounce.  At later times (lower neutrino luminosities), even lower magnetic field strengths $\lesssim 10^{13}$ G similar to those of normal radio pulsars are sufficient for force-free conditions.}
\item{The force-free assumption breaks down in the closed zone due to the high pressure of the co-rotating, hydrostatic atmosphere.  This is estimated to increase the fraction of the surface open to outflows (Y point radius $R_{\rm Y} \sim R_{\rm L}/3$ for $B_0 = 10^{15}$ G; Fig.~\ref{fig:hydrostatic}) as compared to the force-free limit $R_{\rm Y} = R_{\rm L}$.}
\item{Magnetized rotating PNS outflows differ from the standard spherical case as a result of the more rapid divergence of the dipolar field geometry and the effects of magneto-centrifugal acceleration.  The latter dominates the influence on the expansion timescale of the outflow $t_{\rm exp}$ for spin periods $P \lesssim 4$ ms.}
\item{The asymptotic value of the electron fraction is substantially reduced below its neutrino equilibrium value (eq.~[\ref{eq:Yeeq}]) only for very short rotation periods, $P \sim 1$ ms, near centrifugal break-up.}
\item{The mass loss rate per surface area is enhanced by centrifugal effects along low latitude field lines (Fig.~\ref{fig:mdottwoms}).  The smaller fraction of the PNS surface open to outflows as compared to the spherical case (eq.~[\ref{eq:fopen}]) however more than compensates, resulting in the total mass loss rate being smaller by a factor $\gtrsim 10$ than the equivalent luminosity spherical case for $P \gtrsim 2$ ms (Fig.~\ref{fig:mdotint}, Tables \ref{table:integrated}, \ref{table:integrated2Msun}).}
\item{Outflows from proto-magnetars with rotation periods $P \sim 2-5$ ms produce conditions more favorable for the $r$-process by a factor of up to $\approx$ 4 in the relevant parameter $\eta = S^{3}/Y_{e}^{3}t_{\rm exp}$ for spin periods $P \sim 2-5$ ms as compared to spherical, non-rotating wind of the same neutrino luminosity (Fig.~ \ref{fig:sctautwoms}, \ref{fig:sctauint}).  This is similar to the largest enhancement found by \citet{Metzger+07} (see their Fig.~11) for equatorial monopole wind, but for somewhat different reasons.}
\item{The reason for the enhancement in $r$-process conditions in our proto-magnetar wind calculations is as follows.  Open field lines leave the PNS surface close to the rotation axis with a sizable vertical component (Fig.~\ref{fig:schematic}).  This implies that centrifugal effects are less important in the heating region than in the case of an equatorial outflow, resulting in an entropy gain which is only moderately reduced from the non-rotating, spherical case.  At larger radii. however, the field lines become more inclined, resulting in greater centrifugal acceleration and a much faster expansion across radii where seed nuclei form.}
\item{The value of $\eta$ for our most promising solutions are only $\sim$ 10 per cent of the required third-peak threshold value (for $Y_e = 0.45$ and $M = 1.4M_{\odot}$) based on the criterion of \citet{Hoffman+97}, although including the factor of $\sim 3-4$ enhancement due to GR brings our calculations to the brink of success.  A full network calculation is necessary to more precisely quantify the nucleosynthetic yields of magnetar birth.}
\item{Magnetars with birth periods $P \gtrsim 4$ ms represent an appealing site for Galactic $r$-process source, consistent with a variety of observational constraints, including (1) the requirement to not overproduce the energies of magnetar hosting supernova remnants; (2) within uncertainties, the mass of $r$-process nuclei per event needed to explain observed galactic abundances (eq.~[\ref{eq:Mej}]); (3) low mass per event, consistent with previous studies of Galactic chemical evolution; (4) the high estimated masses of the progenitor stars of magnetar imply higher PNS masses, which favor the $r$-process (Tables \ref{table:2Msun}, \ref{table:integrated2Msun}); (5) recent evidence for the birth of millisecond magnetars in low metallicity environments based on the discovery of superluminous supernovae.}   
\item{Proto-magnetar outflows with spin periods $P \sim 1-2$ ms in the range necessary to explain the energetics of GRBs come naturally loaded with the necessary baryon flux to achieve the observed Lorentz factors $\Gamma \sim 100-1000$ of GRB jets (Fig.~\ref{fig:gamma}).}
\item{Magnetar powered GRB jets may be composed of heavy nuclei and helium instead of protons due to their direct synthesis in the outflow.  This could explain the puzzling discovery by the Pierre Auger Observatory that UHECRs are dominated by heavy nuclei at the highest cosmic ray energies.}
\end{itemize}
\section*{Acknowledgments}
We thank Andrey Timokhin for supplying us with the geometry of the force-free magnetosphere; Rodrigo Fernandez for technical help and for providing us with neutrino heating and cooling rates; and Frank Timmes for providing open source codes used in this project. TAT thanks Ondrej Pejcha for stimulating discussions.
\appendix
\section{Results for Massive PNS}
\label{sec:appendix}
\begin{table*}
\begin{scriptsize}
\begin{center}
\vspace{0.05 in}\caption{Summary of Wind Models for $R_{\rm ns}=12$ km, $M=2 M_{\odot}$}
\label{table:2Msun}
\begin{tabular}{ccccccccccc}
\hline \hline
\multicolumn{1}{c}{$P$} &
\multicolumn{1}{c}{$L_{\nu}^{(a)}$} &
\multicolumn{1}{c}{$\frac{\theta}{\theta_{\rm max}}$} & 
\multicolumn{1}{c}{$\dot{M}^{(b)}$} & 
\multicolumn{1}{c}{$S^{(c)}$} &
\multicolumn{1}{c}{$t_{\rm exp}^{(d)}$} &
\multicolumn{1}{c}{$Y_e^{(e)}$} &
\multicolumn{1}{c}{$\eta = S^3/t_{\rm exp} Y_e^3$} &
\multicolumn{1}{c}{$R_{0.5\rm MeV}^{(f)}$} & 
\multicolumn{1}{c}{$\frac{P_{\rm gas}}{P_{\rm mag}}|_{R_{0.5\rm MeV}}^{(g)}$}\\
\hline
 (ms) & (10$^{51}$ erg s$^{-1}$)  &  - & ($M_{\odot}$ s$^{-1}$) & (k$_{\rm b}$ n$^{-1}$) & (ms) & - & $\%$ of ($\eta_{\rm thr}^{(h)}$) & (km) & -\\
\hline 
\\
$^{*}$N/A & 1 & $^{*}$N/A & $1.6\cdot 10^{-6}$ & 144 & $236.4$ & 0.4539 & 1.7 & 45 & $^{*}$N/A \\
$^{*}$N/A & 3 & $^{*}$N/A & $3.0\cdot 10^{-5}$ & 111 & $51.0$ & 0.4531 & 3.6 & 63 & $^{*}$N/A \\
$^{*}$N/A & 6 & $^{*}$N/A & $1.8\cdot 10^{-4}$ & 95 & $22.6$ & 0.4529 & 5.2 & 81 & $^{*}$N/A \\
$^{*}$N/A & 10 & $^{*}$N/A & $7.0\cdot 10^{-4}$ & 86 & $13.7$ & 0.4526 & 6.3 & 101 & $^{*}$N/A \\


$2^\dagger$ & 10 & 0.7 & $1.2\cdot 10^{-4}$ & 135 & 1.6 & 0.4525 & 214.8 & 31 & $1.0\cdot 10^{-2}$\\

2 & 1 & $0.0$ & $2.4\cdot 10^{-6}$ & 139 & 185.2 & 0.4538 & $2.0$ & 35 & $1.6\cdot 10^{-2}$\\
2 & 3 & $0.0$ & $4.4\cdot 10^{-5}$ & 109 & 36.9 & 0.4529 & $4.8$ & 46 & $7.7\cdot 10^{-2}$\\
2 & 6 & $0.0$ & $2.8\cdot 10^{-4}$ & 95 & 14.7 & 0.4526 & $8.0$ & 55 & $2.3\cdot 10^{-1}$\\
2 & 10 & $0.0$ & $1.1\cdot 10^{-3}$ & 87 & 8.2 & 0.4523 & $11.0$ & 64 & $5.7\cdot 10^{-1}$\\
2 & 1 & $0.3$ & $2.8\cdot 10^{-6}$ & 103 & 57.0 & 0.4539 & $2.6$ & 28 & $4.9\cdot 10^{-3}$\\
2 & 10 & $0.3$ & $1.3\cdot 10^{-3}$ & 66 & 1.8 & 0.4519 & $22.4$ & 46 & $1.2\cdot 10^{-1}$\\
2 & 1 & $0.7$ & $3.4\cdot 10^{-6}$ & 86 & 29.3 & 0.4539 & $3.0$ & 25 & $2.8\cdot 10^{-3}$\\
2 & 3 & $0.7$ & $6.4\cdot 10^{-5}$ & 68 & 4.2 & 0.4525 & $10.5$ & 29 & $8.7\cdot 10^{-3}$\\
2 & 6 & $0.7$ & $4.0\cdot 10^{-4}$ & 61 & 1.6 & 0.4518 & $19.1$ & 34 & $2.5\cdot 10^{-2}$\\
2 & 10 & $0.7$ & $1.5\cdot 10^{-3}$ & 57 & 1.3 & 0.4513 & $19.5$ & 39 & $7.7\cdot 10^{-2}$\\
2 & 10 & $1.0$ & $1.9\cdot 10^{-3}$ & 51 & 0.7 & 0.4504 & $25.9$ & 36 & $8.0\cdot 10^{-2}$\\
2 & 6 & $1.0$ & $5.0\cdot 10^{-4}$ & 53 & 1.0 & 0.4511 & $20.4$ & 30 & $2.1\cdot 10^{-2}$\\
3 & 1 & $0.1$ & $2.4\cdot 10^{-6}$ & 139 & 186.7 & 0.4538 & $2.0$ & 35 & $7.5\cdot 10^{-3}$\\
3 & 3 & $0.1$ & $4.4\cdot 10^{-5}$ & 110 & 37.1 & 0.4529 & $4.9$ & 46 & $3.5\cdot 10^{-2}$\\
3 & 6 & $0.1$ & $2.8\cdot 10^{-4}$ & 96 & 14.7 & 0.4526 & $8.2$ & 55 & $1.0\cdot 10^{-1}$\\
3 & 1 & $0.3$ & $2.6\cdot 10^{-6}$ & 115 & 87.6 & 0.4539 & $2.4$ & 30 & $3.4\cdot 10^{-3}$\\
3 & 3 & $0.3$ & $4.9\cdot 10^{-5}$ & 90 & 14.3 & 0.4528 & $7.1$ & 38 & $1.2\cdot 10^{-2}$\\
3 & 6 & $0.3$ & $3.1\cdot 10^{-4}$ & 79 & 5.3 & 0.4524 & $13.0$ & 44 & $3.3\cdot 10^{-2}$\\
3 & 10 & $0.3$ & $1.2\cdot 10^{-3}$ & 73 & 2.9 & 0.4521 & $18.5$ & 51 & $8.6\cdot 10^{-2}$\\
3 & 1 & $0.7$ & $2.9\cdot 10^{-6}$ & 101 & 52.5 & 0.4539 & $2.7$ & 28 & $2.1\cdot 10^{-3}$\\
3 & 3 & $0.7$ & $5.5\cdot 10^{-5}$ & 80 & 7.9 & 0.4527 & $8.8$ & 33 & $6.9\cdot 10^{-3}$\\
3 & 10 & $0.7$ & $1.3\cdot 10^{-3}$ & 66 & 1.7 & 0.4518 & $23.0$ & 45 & $5.3\cdot 10^{-2}$\\
4 & 1 & $0.1$ & $2.4\cdot 10^{-6}$ & 140 & 187.5 & 0.4538 & $2.0$ & 35 & $4.2\cdot 10^{-3}$\\
4 & 3 & $0.1$ & $4.4\cdot 10^{-5}$ & 110 & 37.3 & 0.4528 & $4.9$ & 46 & $2.0\cdot 10^{-2}$\\
4 & 6 & $0.1$ & $2.8\cdot 10^{-4}$ & 96 & 14.7 & 0.4525 & $8.2$ & 55 & $6.0\cdot 10^{-2}$\\
4 & 10 & $0.1$ & $1.1\cdot 10^{-3}$ & 88 & 8.2 & 0.4523 & $11.4$ & 64 & $1.5\cdot 10^{-1}$\\
4 & 1 & $0.4$ & $2.6\cdot 10^{-6}$ & 120 & 105.1 & 0.4539 & $2.2$ & 31 & $2.3\cdot 10^{-3}$\\
4 & 6 & $0.4$ & $3.0\cdot 10^{-4}$ & 82 & 6.6 & 0.4525 & $11.7$ & 46 & $2.4\cdot 10^{-2}$\\
4 & 10 & $0.4$ & $1.1\cdot 10^{-3}$ & 76 & 3.6 & 0.4522 & $16.5$ & 54 & $6.2\cdot 10^{-2}$\\
4 & 1 & $0.7$ & $2.8\cdot 10^{-6}$ & 108 & 67.8 & 0.4539 & $2.5$ & 29 & $1.5\cdot 10^{-3}$\\
4 & 6 & $0.7$ & $3.2\cdot 10^{-4}$ & 74 & 3.9 & 0.4525 & $14.6$ & 41 & $1.4\cdot 10^{-2}$\\
4 & 10 & $0.7$ & $1.2\cdot 10^{-3}$ & 69 & 2.2 & 0.4521 & $20.8$ & 47 & $3.8\cdot 10^{-2}$\\
5 & 1 & $0.1$ & $2.3\cdot 10^{-6}$ & 140 & 187.8 & 0.4538 & $2.0$ & 35 & $2.7\cdot 10^{-3}$\\
5 & 6 & $0.1$ & $2.7\cdot 10^{-4}$ & 96 & 14.8 & 0.4525 & $8.2$ & 55 & $3.9\cdot 10^{-2}$\\
5 & 10 & $0.1$ & $1.1\cdot 10^{-3}$ & 88 & 8.2 & 0.4522 & $11.4$ & 64 & $9.7\cdot 10^{-2}$\\
5 & 1 & $0.4$ & $2.5\cdot 10^{-6}$ & 126 & 127.8 & 0.4538 & $2.1$ & 33 & $1.8\cdot 10^{-3}$\\
5 & 3 & $0.4$ & $4.6\cdot 10^{-5}$ & 99 & 22.8 & 0.4529 & $5.9$ & 41 & $7.5\cdot 10^{-3}$\\
5 & 6 & $0.4$ & $2.9\cdot 10^{-4}$ & 87 & 8.6 & 0.4525 & $10.4$ & 49 & $2.1\cdot 10^{-2}$\\
5 & 10 & $0.4$ & $1.1\cdot 10^{-3}$ & 80 & 4.7 & 0.4523 & $14.7$ & 57 & $5.3\cdot 10^{-2}$\\
5 & 1 & $0.6$ & $2.6\cdot 10^{-6}$ & 116 & 90.4 & 0.4539 & $2.3$ & 31 & $1.3\cdot 10^{-3}$\\
5 & 3 & $0.6$ & $4.8\cdot 10^{-5}$ & 91 & 14.8 & 0.4529 & $6.9$ & 38 & $4.9\cdot 10^{-3}$\\
5 & 10 & $0.6$ & $1.2\cdot 10^{-3}$ & 73 & 3.1 & 0.4522 & $17.9$ & 51 & $3.4\cdot 10^{-2}$\\
10 & 3 & $0.1$ & $4.3\cdot 10^{-5}$ & 109 & 36.4 & 0.4528 & $4.9$ & 46 & $3.3\cdot 10^{-3}$\\
10 & 6 & $0.1$ & $2.7\cdot 10^{-4}$ & 95 & 14.4 & 0.4525 & $8.2$ & 55 & $9.8\cdot 10^{-3}$\\
10 & 10 & $0.1$ & $1.0\cdot 10^{-3}$ & 87 & 8.0 & 0.4522 & $11.5$ & 64 & $2.5\cdot 10^{-2}$\\
10 & 1 & $0.4$ & $2.4\cdot 10^{-6}$ & 136 & 167.2 & 0.4538 & $2.0$ & 34 & $6.3\cdot 10^{-4}$\\
10 & 3 & $0.4$ & $4.4\cdot 10^{-5}$ & 107 & 32.3 & 0.4528 & $5.1$ & 44 & $2.9\cdot 10^{-3}$\\
10 & 6 & $0.4$ & $2.8\cdot 10^{-4}$ & 93 & 12.6 & 0.4525 & $8.8$ & 53 & $8.4\cdot 10^{-3}$\\
10 & 10 & $0.4$ & $1.1\cdot 10^{-3}$ & 85 & 6.9 & 0.4522 & $12.3$ & 62 & $2.1\cdot 10^{-2}$\\
10 & 1 & $0.6$ & $2.5\cdot 10^{-6}$ & 132 & 150.4 & 0.4538 & $2.1$ & 34 & $5.7\cdot 10^{-4}$\\
10 & 3 & $0.6$ & $4.6\cdot 10^{-5}$ & 104 & 28.2 & 0.4528 & $5.4$ & 43 & $2.5\cdot 10^{-3}$\\
10 & 6 & $0.6$ & $2.9\cdot 10^{-4}$ & 90 & 10.8 & 0.4525 & $9.4$ & 51 & $7.2\cdot 10^{-3}$\\
1 & 1 & $0.0$ & $2.4\cdot 10^{-6}$ & 135 & 172.3 & 0.4539 & $1.9$ & 35 & $5.7\cdot 10^{-2}$\\
1 & 3 & $0.0$ & $4.4\cdot 10^{-5}$ & 106 & 33.5 & 0.4528 & $4.8$ & 46 & $2.5\cdot 10^{-1}$\\
1 & 6 & $0.0$ & $2.8\cdot 10^{-4}$ & 92 & 13.3 & 0.4524 & $7.9$ & 55 & $7.0\cdot 10^{-1}$\\
1 & 10 & $0.0$ & $1.1\cdot 10^{-3}$ & 83 & 7.7 & 0.4522 & $10.4$ & 66 & $1.7\cdot 10^{0}$\\
1 & 1 & $0.4$ & $4.2\cdot 10^{-6}$ & 71 & 16.1 & 0.4537 & $3.1$ & 23 & $6.8\cdot 10^{-3}$\\
1 & 6 & $0.4$ & $4.9\cdot 10^{-4}$ & 50 & 1.0 & 0.4506 & $18.4$ & 31 & $6.8\cdot 10^{-2}$\\
1 & 10 & $0.4$ & $1.9\cdot 10^{-3}$ & 47 & 0.7 & 0.4497 & $22.5$ & 36 & $2.7\cdot 10^{-1}$\\
1 & 10 & $0.7$ & $3.9\cdot 10^{-3}$ & 33 & 0.4 & 0.4416 & $14.9$ & 31 & $4.7\cdot 10^{-1}$\\
1 & 1 & $1.0$ & $2.0\cdot 10^{-5}$ & 31 & 2.3 & 0.4416 & $2.0$ & 17 & $7.4\cdot 10^{-3}$\\
1 & 6 & $1.0$ & $2.7\cdot 10^{-3}$ & 24 & 0.3 & 0.4224 & $8.5$ & 24 & $2.2\cdot 10^{-1}$\\
1 & 10 & $1.0$ & $1.0\cdot 10^{-2}$ & 24 & 0.2 & 0.4200 & $10.3$ & 29 & $1.6\cdot 10^{0}$\\

\\
\hline
\hline
\end{tabular}
\end{center}
$^{*}$Spherical solutions; $^{\dagger}$ Calculated using Pacynzsky-Wiita gravitational potential to approximate effects of GR; $^{(a)}$ $L_{\nu} \equiv (L_{\nu_e} + L_{\bar{\nu}_e})/2$; $^{(b)}$ Mass outflow rate (spherical, or equivalent spherical; eq.~[\ref{eq:Mdot}]); $^{(c)}$Final entropy $^{(d)}$Expansion timescale at $T = 0.5$ MeV (eq.~[\ref{eq:texp}]), approximate location of alpha particle formation; $^{(e)}$Final electron fraction; $^{(f)}$Radius at which $T = 0.5$ MeV; $^{(g)}$Ratio of `matter pressure' $P_{\rm gas}$ to magnetic pressure $P_{\rm mag} = B^{2}/8\pi$ (for an assumed surface field $B_0 = 10^{15}$ G) at $T = 0.5$ MeV, where $P_{gas}$ includes gas pressure $P$, thermal energy density $e$, kinetic energy density (rest energy subtracted); $^{(h)}$Threshold ratio for outflow to attain third-peak $r$-process (eq.~[\ref{eq:eta}]);  
\end{scriptsize}
\end{table*}

\begin{table*}
\begin{scriptsize}
\begin{center}
\vspace{0.05 in}\caption{Summary of Integrated Quantities for $R_{NS}=12$ km, $M=2 M_{\odot}$}
\label{table:integrated2Msun}
\begin{tabular}{cccccc}
\hline \hline
\multicolumn{1}{c}{$P$} &
\multicolumn{1}{c}{$L_{\nu}^{(a)}$} &
\multicolumn{1}{c}{$\dot{M}_{\rm tot}^{(b)}$} & 
\multicolumn{1}{c}{$M_{\rm ej}^{(c)}$} &
\multicolumn{1}{c}{$\Gamma_{\rm max}^{(d)}$} &
\multicolumn{1}{c}{$\langle \eta \rangle = \langle S^{3}/t_{\rm exp}Y_{e}^{3}\rangle^{(e)}$} \\
\hline
 (ms) & (10$^{51}$ erg s$^{-1}$) &  ($M_{\odot}$ s$^{-1}$) & ($M_{\odot}$) & - & ($\%$ of $\eta_{\rm thr}$) \\
\hline 
\\\
$^{*}$N/A & 1 & $1.6\cdot 10^{-6}$ & $1.5\cdot 10^{-5}$ & $^{*}$N/A & 1.7 \\
$^{*}$N/A & 3 & $3.0\cdot 10^{-5}$ & $1.1\cdot 10^{-4}$ & $^{*}$N/A & 3.6 \\
$^{*}$N/A & 6 & $1.8\cdot 10^{-4}$ & $3.8\cdot 10^{-4}$ & $^{*}$N/A & 5.2 \\
$^{*}$N/A & 10 & $7.0\cdot 10^{-4}$ & $1.3\cdot 10^{-3}$ & $^{*}$N/A & 6.3 \\

1 & 1 & $1.5\cdot 10^{-6}$ & $1.4\cdot 10^{-5}$ & 580.3 & 2.6 \\
1 & 10 & $6.6\cdot 10^{-4}$ & $1.2\cdot 10^{-3}$ & 2.3 & 14.2 \\
1 & 3 & $6.1\cdot 10^{-6}$ & $2.3\cdot 10^{-5}$ & 143.5 & 4.8 \\
1 & 6 & $2.0\cdot 10^{-4}$ & $4.1\cdot 10^{-4}$ & 5.4 & 13.7 \\
2 & 1 & $2.1\cdot 10^{-7}$ & $1.9\cdot 10^{-6}$ & 261.7 & 2.9 \\
2 & 10 & $9.9\cdot 10^{-5}$ & $1.9\cdot 10^{-4}$ & 1.5 & 31.8 \\
2 & 3 & $3.8\cdot 10^{-6}$ & $1.4\cdot 10^{-5}$ & 15.2 & 9.3 \\
2 & 6 & $2.6\cdot 10^{-5}$ & $5.4\cdot 10^{-5}$ & 3.1 & 17.3 \\
3 & 1 & $1.4\cdot 10^{-7}$ & $1.3\cdot 10^{-6}$ & 78.4 & 2.6 \\
3 & 10 & $6.2\cdot 10^{-5}$ & $1.2\cdot 10^{-4}$ & 1.2 & 21.8 \\
3 & 3 & $2.6\cdot 10^{-6}$ & $9.6\cdot 10^{-6}$ & 5.2 & 8.2 \\
3 & 6 & $1.5\cdot 10^{-5}$ & $3.1\cdot 10^{-5}$ & 1.7 & 12.7 \\
4 & 1 & $1.1\cdot 10^{-7}$ & $9.8\cdot 10^{-7}$ & 32.8 & 2.4 \\
4 & 10 & $4.8\cdot 10^{-5}$ & $8.9\cdot 10^{-5}$ & 1.1 & 19.0 \\
4 & 3 & $1.7\cdot 10^{-6}$ & $6.4\cdot 10^{-6}$ & 3.0 & 4.9 \\
4 & 6 & $1.2\cdot 10^{-5}$ & $2.6\cdot 10^{-5}$ & 1.3 & 13.4 \\
5 & 1 & $9.1\cdot 10^{-8}$ & $8.4\cdot 10^{-7}$ & 16.3 & 2.3 \\
5 & 10 & $4.1\cdot 10^{-5}$ & $7.6\cdot 10^{-5}$ & 1.0 & 16.9 \\
5 & 3 & $1.7\cdot 10^{-6}$ & $6.3\cdot 10^{-6}$ & 1.8 & 6.7 \\
5 & 6 & $1.0\cdot 10^{-5}$ & $2.1\cdot 10^{-5}$ & 1.1 & 10.3 \\
10 & 1 & $3.6\cdot 10^{-8}$ & $3.3\cdot 10^{-7}$ & 3.4 & 2.0 \\
10 & 10 & $1.6\cdot 10^{-5}$ & $3.0\cdot 10^{-5}$ & 1.0 & 12.3 \\
10 & 3 & $6.7\cdot 10^{-7}$ & $2.5\cdot 10^{-6}$ & 1.1 & 5.3 \\
10 & 6 & $4.2\cdot 10^{-6}$ & $8.9\cdot 10^{-6}$ & 1.0 & 9.2 \\

 \\
\hline
\hline
\end{tabular}
\end{center}
$^{*}$Spherical Solutions. 
$^{(a)}$ $L_{\nu} \equiv (L_{\nu_e} + L_{\bar{\nu}_e})/2$; $^{(b)}$Total mass loss rate, integrated across the open zone $\theta \in [0,\theta_{\rm max}]$.  $^{(c)}$Total ejected mass at luminosities $\sim L_{\nu}$, estimated using the PNS cooling evolution from \citet{Pons+99}.  $^{(d)}$Maximum Lorentz factor achieved by the proto-magnetar outflow, calculated assuming a surface field $B_{0} = 3\times 10^{15}$ G. $^{(e)}$Threshold for $r$-process (eq.~[\ref{eq:eta}]), in a mass-weighted averaged over the open zone from $\theta \in [0,\theta_{\rm max}]$.
\end{scriptsize}
\end{table*}


\end{document}